\documentclass[12pt]{article}
\usepackage{tikz}
\usetikzlibrary{matrix,arrows,decorations.pathmorphing,decorations.pathreplacing}
\usepackage{jheppub}
\usepackage{mathpazo}
\usepackage{amsmath,amssymb,euscript,array,mathrsfs,appendix,ctable,marvosym}
\usepackage{mathtools}
\usepackage{arydshln}
\usepackage{todonotes}
\usepackage{graphicx}

\newenvironment{changemargin}[2]{%
  \begin{list}{}{%
    \setlength{\topsep}{0pt}%
    \setlength{\leftmargin}{#1}%
    \setlength{\rightmargin}{#2}%
    \setlength{\listparindent}{\parindent}%
    \setlength{\itemindent}{\parindent}%
    \setlength{\parsep}{\parskip}%
  }%
  \item[]}{\end{list}}

\def\no{\noindent}

\def\EA{{\EuScript A}}
\def\MS{{\EuScript M}}
\def\EE{{\EuScript E}}

\def\BH{{\cal H}}

\newcommand{\Tr}{\operatorname{Tr}}
\newcommand{\IM}{\operatorname{Im}}

\def\bra#1{\langle #1|}
\def\ket#1{| #1\rangle}

\def\LCATH{\raisebox{-2pt}{\begin{tikzpicture}[scale=0.14]
\draw (0,0) ellipse (1.8cm and 1.2cm);
\draw [-] (1,0) -- (2.3,0);
\draw [-] (-1,0) -- (-2.3,0);
\draw [-] (1,-0.2) -- (2.2,-0.5);
\draw [-] (-1,-0.2) -- (-2.2,-0.5);
\draw [-] (1,0.2) -- (2.2,0.5);
\draw [-] (-1,0.2) -- (-2.2,0.5);
\filldraw[black] (0.4,0.3) circle (0.1cm);
\filldraw[black] (-0.4,0.3) circle (0.1cm);
\draw[-] (-0.5,-0.6) to[out=-30,in=210] (0.5,-0.6);
\draw[-] (0.8,1.1) -- (1.1,1.8) -- (1.4,0.8);
\draw[-] (-0.8,1.1) -- (-1.1,1.8) -- (-1.4,0.8);
\end{tikzpicture}
}\!}

\def\DCATH{\raisebox{-2pt}{\begin{tikzpicture}[scale=0.14]
\draw (0,0) ellipse (1.8cm and 1.2cm);
\draw [-] (1,0) -- (2.3,-0.5);
\draw [-] (-1,0) -- (-2.3,-0.5);
\draw [-] (1,-0.2) -- (2.2,-1);
\draw [-] (-1,-0.2) -- (-2.2,-1);
\draw [-] (1,0.2) -- (2.2,0);
\draw [-] (-1,0.2) -- (-2.2,0);
\draw[-] (0.2,0.1) -- (0.6,0.5);
\draw[-] (0.2,0.5) -- (0.6,0.1);
\draw[-] (-0.2,0.1) -- (-0.6,0.5);
\draw[-] (-0.2,0.5) -- (-0.6,0.1);
\draw[-]  (-0.5,-0.6) to[out=30,in=150] (0.5,-0.6);
\draw[-] (0.8,1.1) -- (1.1,0.6) -- (1.4,0.8);
\draw[-] (-0.8,1.1) -- (-1.1,0.6) -- (-1.4,0.8);
\end{tikzpicture}
}\!}

\newcommand{\MAT}[1]{\begin{pmatrix} #1\end{pmatrix}}

\newcommand{\EQ}[1]{\begin{equation}\begin{split} #1
\end{split}\end{equation}}

\newcounter{Part}

\title{The Copenhagen Interpretation Born Again}
\author{Timothy J. Hollowood}
\affiliation{
Department of Physics, Swansea University,\\ Swansea, SA2 8PP, United Kingdom
}

\emailAdd{t.hollowood@swansea.ac.uk}

\abstract{An approach to quantum mechanics is developed
which makes the Heisenberg cut between the deterministic microscopic quantum world and the partly deterministic, partly stochastic macroscopic world explicit. The microscopic system evolves according to the Schr\"odinger equation with  
stochastic behaviour arising when the system is probed by a set of coarse grained macroscopic observables whose resolution scale defines the Heisenberg cut. The resulting stochastic process can account for the different facets of the classical limit: Newton's laws (ergodicity broken); statistical mechanics of thermal ensembles (ergodic); and solve the measurement problem (partial ergodicity breaking).  In particular, the usual rules of the Copenhagen interpretation, like the Born rule, emerge, along with completely local descriptions of EPR type
experiments. The formalism also re-introduces a dynamical picture of equilibration and thermalization in quantum statistical mechanics and 
provides insight into how classical statistical mechanics can arise in the classical limit and in a way that alleviates various conceptual problems. 
}

\notoc
\begin{document}

\maketitle

\newpage

\section{Introduction}\label{s1}

Quantum mechanics is an amazingly successful and universal framework that describes physics over a huge range of scales, from the quantum theory of gravity provided by fundamental strings, through particle physics, nuclear physics, atomic physics to condensed matter physics. There can be little doubt that we know how to {\it interpret\/} quantum mechanics from a practical point of view by employing the rules of the Copenhagen interpretation.\footnote{It can be argued that there is no single, historically coherent version of the Copenhagen interpretation. However, the rules are well known to, and used by, most physicists and also form the basis of most textbook treatments of quantum mechanics.}

The Copenhagen interpretation has all the tools that are needed to unlock the microscopic quantum world in the laboratory. It works by hypothesising the existence of the Heisenberg cut, a division between the deterministic microscopic quantum world governed by the Schr\"odinger equation
 and the classical, partly deterministic, partly stochastic, macroscopic world governed by classical mechanics and the Born rule. It is worth amplifying this point: it is not the microscopic world that is probabilistic and the classical world deterministic, the reality is the opposite as is apparent when one listens to a Geiger counter. Moreover, 
the classical world includes statistical mechanics and hence is stochastic even if the r\^ole of probability is not conceptually well understood in that context.
It is important that the scale of the Heisenberg cut can be changed somewhat without altering the macroscopic predictions as long as it is kept below (as a scale in position space) macroscopic scales.
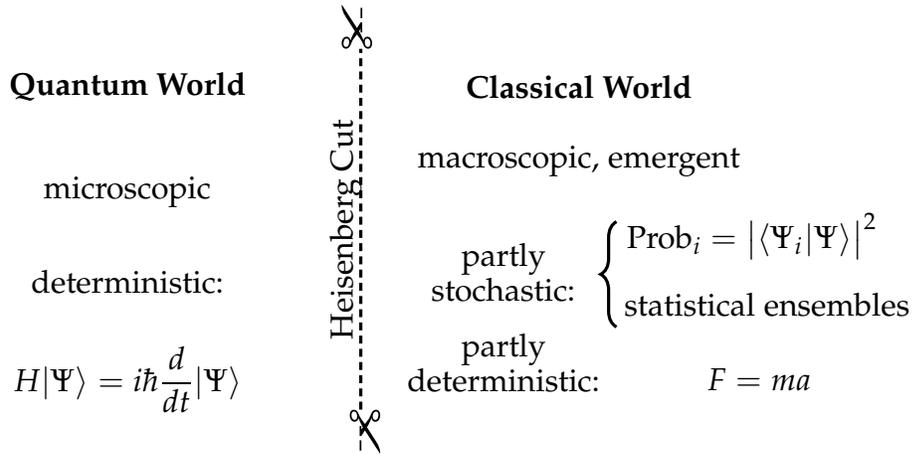
\begin{center}
\begin{tikzpicture}[scale=1]
\draw[densely dashed,very thick] (3.1,-2.3) -- (3.1,2.5);
\node at (0,2) (a1) {\bf Quantum World};
\node at (6,2) (a2) {\bf Classical World};
\node at (0,0.6) (a1) {microscopic};
\node at (6,1) (a2) {macroscopic, emergent};
\node at (0,-0.6) (a1) {deterministic:};
\node at (5,-0.7) (a2) {stochastic:};
\node at (5,-0.3) (a2) {partly};
\node at (5,-1.9) (a2) {deterministic:};
\node at (5,-1.5) (a2) {partly};
\node at (8.4,-1.9) (a2) {$F=ma$};
\node at (0,-1.9) (a1) {$H\ket{\Psi}=i\hbar\dfrac{d}{dt}\ket{\Psi}$};
\node at (8.3,-0) (a2) {$\text{Prob}_i=\big|\bra{\Psi_i}\Psi\rangle\big|^2$};
\node at (8.5,-0.9) (a2) {statistical ensembles};
\node[rotate=90] at (2.9,0.1) (a1) {Heisenberg Cut};
\node[rotate=-90] at (3.1382,-2.6) {\Huge\Cutright};
\node[rotate=90] at (3.066,2.8) {\Huge\Cutright};
\node at (-3,0) (k1) {\phantom{.}};
\draw[decoration={brace,amplitude=0.5em},decorate,very thick] (6.5,-1.2) -- (6.5,0.2);
\end{tikzpicture}
\end{center}

The Copenhagen interpretation posits a kind of stochastic dynamics that kicks in whenever a measurement is made on a microscopic quantum system. So during a measurement process where an initial state vector $\ket{\Psi(0)}$ evolves into the linear combination 
\EQ{
\ket{\Psi(T)}=\sum_i c_i\ket{\Psi_i}\ ,
}
and where the component states $\ket{\Psi_i}$ are macroscopically distinct,
the final state is then interpreted as describing an ensemble with probabilities are given by the Born rule
\EQ{
\text{prob}_i=|c_i|^2\ .
}
This begs some questions of which two key ones are:

\begin{enumerate}
\item When does the underlying pure state become interpreted as an ensemble, i.e.~where is the Heisenberg cut to be placed?
\item Why are the macroscopically distinct states $\ket{\Psi_i}$ picked out as special?
\end{enumerate}

The conventional answer to (i) is that it doesn't matter where we put the Heisenberg cut as long as it is somewhere between the microscopic and macroscopic while (ii) is solved by fiat. The Copenhagen interpretation in its present form has shown itself
sufficient to use quantum mechanics in practice. Is there, therefore, any merit in trying to go beyond the conventional rules in order to provide more convincing answers to the questions above? The danger is that this might add to the feeling that the Copenhagen interpretation is inadequate which is not our intention. On the other hand, as we will try to argue, there is the potential to unify both the deterministic and stochastic aspects of the classical world, the latter including both quantum measurement scenarios as well as statistical mechanics (SM).

The enhanced formalism we are after will determine what are the possible outcomes of a quantum measurement without having to resort to identifying them by hand. In addition, it will provide a better understanding of the Heisenberg cut and therefore how the classical world  emerges from quantum mechanics. In this regard, the existence of a classical world is not guaranteed by the conventional classical limit and the correspondence principle alone. It will also yields a formalism where the reduction of the state vector is just a non-compulsory piece of book keeping.

In addition, an important element of the proposal will be to provide a new way to understand how classical SM can emerge from an underlying quantum system.
It may come as a surprise to learn that classical SM still has many unresolved conceptual problems;\footnote{Wallace \cite{wallace} has a nice review of these conceptual difficulties and how they are related to those of quantum mechanics. Some of the outstanding conceptual issues of classical SM set in a historical context are discussed in the excellent review of Uffink \cite{Uffink}.} including:

\begin{enumerate}
\item The conceptual status of probability in a deterministic theory.
\item Definition of entropy and the second law.
\item Description of non-equilibrium and the approach to equilibrium.
\item The r\^ole of ergodicity and the relation between ensembles and long, or infinite, time averages.
\end{enumerate}

The idea of building SM directly on quantum mechanics has its genesis in the work of 
Schr\"odinger \cite{Sch} and von~Neumann \cite{vN,vNb} back in the 1920s. This kind of approach, taken up (or re-invented) in the modern era,\footnote{For a selection of papers see \cite{PopescuShortWinter:2005fsmeisa,PopescuShortWinter:2006efsm,GoldsteinLebowitzTumulkaZanghi:2006ct,LPSW,Sh,GMM,Reimann1,Reimann2,Gogolin,Srednicki:1995pt,Srednicki2,GLMTZ,RS} and references therein.} misses out classical mechanics completely, and thereby avoids its conceptual problems. But we would surely be missing something by not understanding how classical SM can arise as a limit of quantum mechanics.
In this regard, it would be wasteful not to put the probabilities of quantum mechanics to good use to provide ab initio probabilities for SM. Wallace \cite{wallace} has pointed out that in this context that there is no need to introduce probabilities twice, firstly for quantum purposes and then for statistical mechanical purposes. We intend to elaborate on this point and then suggest that 
if we can understand the stochastic element of quantum mechanics more formally then we may at the same time provide a proper foundation for classical SM as the classical limit of quantum mechanics. There will be a considerable pay off because, as we will see, issues to do with probabilities, entropy, ensembles and ergodicity potentially become simpler in the quantum context. 

With this motivation, we will construct a framework for quantum mechanics for which the following are true:

\begin{enumerate}
\item The underlying system is described by a state vector evolving according to the Schr\"odin-ger equation.
\item The r\^ole of observers is played by sets of observables that describe the macroscopic interactions of a macro-system.\footnote{There is no need for Alices, Bobs, agents, brains, minds, consciousness or users---human, feline or otherwise.}
\item A classical regime and the usual Copenhagen rules of measurement emerge at macroscopic scales: definite macroscopically distinct outcomes are predicted with probabilities that are captured by the Born rule if the measuring device is accurate.
\item The classical limit for macroscopic systems displays both
deterministic and stochastic features which can explain why macroscopic objects have definite positions and obey Newton's laws whilst, at the same time, their internal degrees of freedom are in thermal equilibrium. 
In the classical limit, a dynamical picture of classical SM emerges but with a stochastic dynamics on a discrete set of states for which issues concerning ergodicity are conceptually simpler to investigate.
\end{enumerate}

The present work is a development of earlier work \cite{Hollowood:2013cbr,Hollowood:2013xfa,Hollowood:2013bja} which themselves are built on some aspects of the class of modal interpretations of quantum mechanics. The key idea is that Heisenberg cut is identified implicitly by the resolution scale of a set of relevant macroscopic degree of freedom of a system. 
In \cite{Hollowood:2013cbr,Hollowood:2013xfa,Hollowood:2013bja} this was done by assuming that the Hilbert space has an explicit factorization 
\EQ{
\BH=\BH_A\otimes\BH_E\ ,
}
where $\BH_A$ describes the accessible, or macroscopically relevant, degrees of freedom and $\BH_E$, the environment, encompasses all the remaining degrees of freedom. The tensor product factorization is therefore a manifestation of the Heisenberg cut. This kind of split is useful in toy models for describing the way the formalism works but in realistic situations we cannot expect such a clean split of the Hilbert space and, in addition, it is difficult to see how the Heisenberg cut could be varied in this context.

The fact that the Heisenberg cut can be varied as long as it is kept at sub-macroscopic scales is very reminiscent of the Wilsonian cut off in quantum field theory (QFT) \cite{Wilson:1983}. 
In the modern era, QFTs are viewed as effective theories that have an explicit finite resolution, or coarse graining, scale known as the Wilsonian cut off.\footnote{Even our most fundamental QFT, the standard model, is only an effective theory which needs an explicit cut off to be well defined. It is much more stringent requirement to ask that a QFT has a continuum limit for which the cut off or resolution scale is taken to zero in position space. This kind of limit can only be taken if there is a UV fixed point of the renormalization group.} They are only valid for describing phenomena at larger distances than this scale. The important point is that phenomena on larger length scales are insensitive to changing the Wilsonian cut off as long as the latter is kept below the scale of the phenomena. The analogy with the Heisenberg cut is intriguing. One feature is worth emphasizing: there is a subjective element in choosing the Wilsonian cut off because it depends on the distance scale of the phenomena that you want to focus on. But this kind of subjectivity is completely natural: we are free to choose the length scale of the phenomena we want to investigate and each scale has its own effective theory. 

The lesson for the Heisenberg cut is that it would be natural to associate it to the resolution scale of a set of observables that are needed to describe phenomena on, in this case, macroscopic scales: macroscopic phenomena require a sub-macroscopic Heisenberg cut.
In the present approach, motivated for the need to describe more realistic systems, and with the notion of an effective QFT in mind, we propose to change the 
focus from tensor product factors and reduced density operators, to the set of relevant---or macroscopically accessible, or coarse grained---observables $\EA=\{{\cal O}_n\}$. These are a set of observables that are coarse grained on sub-macroscopic scales that are needed to faithfully describe the 
macroscopic interactions of the system with other macroscopic systems. Implicit in the definition of the set $\EA$, therefore, is the resolution or coarse graining scale which is nothing other than the Heisenberg cut. 
We will find the original tensor product 
and reduced density operator formalism will emerge as a special case of the more general formalism that we develop. In this sense, the present approach subsumes the earlier work \cite{Hollowood:2013cbr,Hollowood:2013xfa,Hollowood:2013bja}.
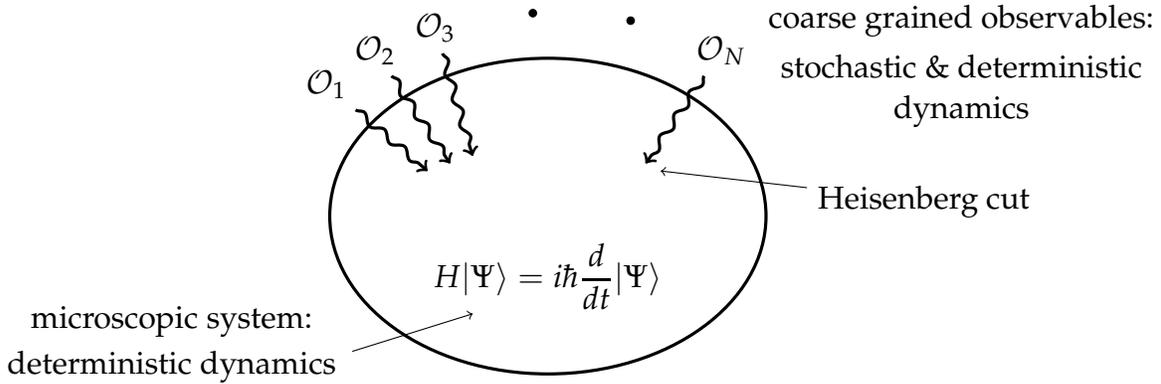
\begin{figure}
\begin{center}
\begin{tikzpicture}[scale=1]
\node at (-2.95,1.75) (a1) {${\cal O}_1$};
\node at (-2.3,2.2) (a2) {${\cal O}_2$};
\node at (-1.5,2.5) (a4) {${\cal O}_3$};
\node at (2.3,2.2) (a3) {${\cal O}_N$};
\filldraw[black] (-0.2,2.7) circle (0.05cm);
\filldraw[black] (1.1,2.6) circle (0.05cm);
\draw[decorate,decoration={snake,amplitude=0.05cm},very thick,->] (a1) -- (-1.6,0.6);
\draw[decorate,decoration={snake,amplitude=0.05cm},very thick,->] (a2) -- (-1.3,0.7);
\draw[decorate,decoration={snake,amplitude=0.05cm},very thick,->] (a4) -- (-1,0.8);
\draw[decorate,decoration={snake,amplitude=0.05cm},very thick,->] (a3) -- (1.3,0.7);
\draw[very thick] (0,0) ellipse  (2.9cm and 2.1cm);
\node at (0,-0.8) (b1) {$H\ket{\Psi}=i\hbar\dfrac{d}{dt}\ket{\Psi}$};
\node at (-5,-1.4) (c1) {microscopic system:};
\node at (-5,-2) (c2) {deterministic dynamics};
\node at (5.5,2.6) (c3) {coarse grained observables:};
\node at (5.5,2) (c4) {stochastic \& deterministic};
\node at (5.5,1.4) (c5) {dynamics};
\node at (5,0.2) (c6) {Heisenberg cut};
\draw[->] (c6) -- (1.5,0.6);
\draw[->] (-2.6,-1.8) -- (-1,-1.3);
\end{tikzpicture}
\end{center}
\caption {\small A quantum system evolves deterministically according to the Schr\"odinger equation. In our proposal, the stochastic element arises when the macroscopic interactions of the system are described by a set of probes in the form of a set of coarse grained observables $\EA=\{{\cal O}_n\}$ whose resolution scale defines the Heisenberg cut.}
\end{figure}

In our approach to the quantum mechanics, the set of coarse grained observables $\EA$ has a perspective on the underlying quantum state with the Heisenberg cut identified with the resolution scale of the observables. Like the conventional Copenhagen interpretation, the 
perspective of $\EA$ involves a kind of stochastic dynamics, but unlike the conventional Copenhagen interpretation there is nothing ad hoc about its definition.
In particular, the Born rule will emerge approximately in realistic situations with efficient measuring devices and state vector reduction will be seen to be a pragmatic---but ultimately unnecessary---piece of book keeping that is performed locally on the state vector. 

It is a key idea of our approach that although the microscopic state of the system is the state vector $\ket{\Psi}$ the latter is not directly the macrostate of a system. This is very natural from the perspective of both QFT and quantum SM:

\begin{enumerate}
\item In QFT the macroscopic, or IR, state of a system is effectively determined by the correlation functions of the low momentum, or coarse grained, operators $\bra{\Psi}{\cal O}_{n_1}\cdots{\cal O}_{n_p}\ket{\Psi}$. 
\item In quantum statistical mechanics, the properties of equilibration and thermalization are 
are properties of the expectation values of coarse grained observables $\bra{\Psi}{\cal O}_n\ket{\Psi}$. In particular, they are  {\it not\/} properties of the underlying state vector that does not equilibrate or thermalize itself.\footnote{It is worth pointing out, and discussed at further length in section \ref{s7}, that the idea of defining the effective state of a system through the expectation values of a set of coarse grained observables plays an important r\^ole in quantum approaches to SM going back to von Neumann \cite{vN}.}
\end{enumerate}

This paper is organized as follows. In section \ref{s2}, we set out the main features of the {\it emergent Copenhagen interpretation\/} along with a limited commentary. Later sections expand upon this discussion beginning in section \ref{s3} which addresses the definition of the quantum microstates in detail. This is followed up in section \ref{s4} by a discussion of the stochastic process that controls the dynamics of the quantum microstates. It is important to show that the stochastic process is local in order to ensure that the formalism is causal. In section \ref{s6} we show how the formalism applies to a simple quantum system consisting of a particle moving in one dimension. Simple though it is, this example draws out some important details. A crude model of measurement is then described in section \ref{s6.5}. This model is not supposed to be realistic but is simple enough to see how the proposal works in practice. In section \ref{s9}, we show how the formalism produces a description of the EPR thought experiment using qubits that involves only local interactions between the qubits and the measuring devices: there is no quantum non-locality.
Section \ref{s7} describes how the formalism has implications for the quantum mechanical underpinning of classical statistic mechanics. Finally, in section \ref{s8}, we draw some conclusions.

\section{The Emergent Copenhagen Interpretation}\label{s2}

In this section, we summarize the three key elements of the emergent Copenhagen interpretation previously introduced in \cite{Hollowood:2013cbr,Hollowood:2013xfa,Hollowood:2013bja} but here generalized. More detail in given in the following sections, particularly \ref{s3} and \ref{s4}. 

\vspace{0.2cm}
\begin{changemargin}{0.3cm}{0.3cm}
\hrule
\vspace{0.2cm}
\no{\bf 1.} {\it The underlying laws of quantum mechanics hold unchanged: in other words, the underlying state $\ket{\Psi}$ evolves according to the Schr\" odinger equation
\EQ{
H\ket{\Psi}=i\hbar\frac{d}{dt}\ket{\Psi}\ .
\label{hem}
}
\hrule}\end{changemargin}

\vspace{0.2cm}
It is possible to relax the condition of the system having a pure state with the generalization to mixed states described in appendix \ref{a1}.

\vspace{0.2cm}
\begin{changemargin}{0.3cm}{0.3cm}
\hrule
\vspace{0.2cm}
\no{\bf 2.} {\it The stochastic element that we observe of quantum system arises 
from the way the underlying quantum system interacts with other systems.
This involves defining a set of probes in the form of coarse grained observables $\EA=\{{\cal O}_n\}$ that describe the potential macroscopic interactions of the system with other systems. 
At time $t$, the state that is actually realized,} from the point of view of the set {\it  $\EA$, i.e.~the macrostate, is encoded in the expectation values $\bra{\Psi_{i(t)}}{\cal O}_n\ket{\Psi_{i(t)}}$, $n=1,2,\ldots,\text{dim}\,\EA$,
where $\ket{\Psi_{i(t)}}$ is
one of the set of {\it quantum microstates\/} $\MS=\{\ket{\Psi_j}\}$ that is realized at time $t$. These are a discrete set of orthonormal states (defined fully in section \ref{s3}) that depend implicitly on both $\ket{\Psi}$ and $\EA$ with the property that the expectation values of operators in the set $\EA$ are effectively captured by an ensemble:
\EQ{
\bra{\Psi}{\cal O}_n\ket{\Psi}=\sum_{i=1}^{\text{dim}\,\MS}|c_i|^2\bra{\Psi_i}{\cal O}_n\ket{\Psi_i}\ ,\qquad\forall\ {\cal O}_n\in\EA\ ,
\label{kww}
}
that has maximal entropy and whose probabilities satisfy the ``microscopic Born rule"
\EQ{
|c_i|^2=\big|\bra{\Psi_i}\Psi\rangle\big|^2\ .
\label{mbr}
}
\hrule}\end{changemargin}

\vspace{0.2cm}
Defining the set of observables $\EA$ and the set of quantum microstates $\MS$ will form the subject of section \ref{s3}. In the simplest case, $\EA$ contains a single observable ${\cal O}$ and then the set $\MS$ consists of the eigenstates of ${\cal O}$ that appear in the expansion of $\ket{\Psi}$. However, in general realistic sets $\EA$ will consist of coarse grained observables in some local region of space that describe the possible macroscopic interactions of the system with other macroscopic systems.

Our approach incorporates the notion of {\it observer complementarity\/} implicit in the choice of the set of observables $\EA$: different sets define different perspectives.\footnote{The fact that perspective plays an important r\^ole in quantum mechanics is implicit in the Copenhagen interpretation: local observers reduce the state vector according to the results of local measurements. It is also important for quantum approaches to statistical mechanics, where phenomena like equilibration and thermalization happen relative to a choice of macroscopic coarse grained observables.} 
The question that naturally arises is whether this introduces a kind of subjectivity into the approach: in other words, is the choice of the set ${\EuScript A}$ determined or is it arbitrary? To add to the discussion in section \ref{s1}, clearly there is some freedom to choose $\EA$, however, this kind of non-uniqueness is 
familiar in physics. If we want to investigate phenomena of a system at a certain length scale the theory should be couched in terms of a set of relevant observables that are determined by the possible probes, or interactions with external systems, at that scale. The choice of scale is subjective but the details of the effective theory relevant at that scale is not. Since we want to investigate how a classical world can emerge from quantum mechanics the relevant scale is macroscopic and $\EA$ are a set of sub-macroscopically coarse grained observables that are relevant for describing macroscopic interactions. There will be some some residual freedom in the choice of $\EA$ and the coarse graining scale---the Heisenberg cut---but we expect that the macroscopic predictions will be insensitive to changing this scale, in direct analogy to the insensitivity of the IR limit of effective QFT to changing the Wilsonian cut off or the way the observables are actually coarse grained.

The expression \eqref{kww} for the expectation values of operators in the set $\EA$ in the state $\ket{\Psi}$ imply that they can be interpreted as a statistical ensemble of the quantum microstates with probabilities $|c_i|^2$. The fact that the probabilities equal $|\bra{\Psi_i}\Psi\rangle|^2$ is a statement of what we call the {\it microscopic Born rule\/}. It will be seen to be crucial to derive the Born rule itself in a measurement scenario.

\vspace{0.2cm}
\begin{changemargin}{0.3cm}{0.3cm}
\hrule
\vspace{0.2cm}
\no{\bf 3.} {\it The actual evolution of the system {\it from the perspective of the set of observables\/} $\EA$ is encoded in the expectation values $\bra{\Psi_{i(t)}}{\cal O}_n\ket{\Psi_{i(t)}}$, $\forall\ {\cal O}_n\in\EA$, for a  sequence of quantum microstates $\ket{\Psi_{i(t)}}$ related by discontinuous transitions; e.g.
\begin{center}
\begin{tikzpicture}[scale=0.6]
\node at (-10.5,1.7) (a1) {$i(t)=$};
\node at (-6.7,3.2) (a2) {$i_2\qquad t<t_1$};
\node at (-6,2.2) (a2) {$i_4\qquad t_1<t<t_2$};
\node at (-6,1.2) (a2) {$i_1\qquad t_2<t<t_3$};
\node at (-6,0.2) (a2) {$\cdots$};
\draw[decoration={brace,amplitude=0.5em},decorate,very thick] (-8.9,0) -- (-8.9,3.5);
\draw[densely dashed] (0,0) -- (8,0);
\draw[densely dashed] (0,1) -- (8,1);
\draw[densely dashed] (0,2) -- (8,2);
\draw[densely dashed] (0,3) -- (8,3);
\node at (-1,0) (a1) {$\ket{\Psi_{i_1}}$};
\node at (-1,1) (a1) {$\ket{\Psi_{i_2}}$};
\node at (-1,2) (a1) {$\ket{\Psi_{i_3}}$};
\node at (-1,3) (a1) {$\ket{\Psi_{i_4}}$};
\node at (2,-0.8) (h1) {$t_1$};
\node at (4,-0.8) (h1) {$t_2$};
\node at (5.5,-0.8) (h1) {$t_3$};
\draw[very thick] (0,1) -- (2,1);
\draw[decorate,decoration={snake,amplitude=0.05cm},very thick,->] (2,1) -- (2,3);
\draw[very thick] (2,3) -- (4,3);
\draw[decorate,decoration={snake,amplitude=0.05cm},very thick,->] (4,3) -- (4,0);
\draw[very thick] (4,0) -- (5.5,0);
\draw[decorate,decoration={snake,amplitude=0.05cm},very thick,->] (5.5,0) -- (5.5,2);
\draw[very thick] (5.5,2) -- (8,2);
\end{tikzpicture}
\end{center}
described by a continuous time Markov chain with transition rates for $\ket{\Psi_j}\to\ket{\Psi_i}$, $i\neq j$,}
\EQ{
T_{ij}=\text{max}\Big(-\frac2\hbar\IM\Big[\frac{c_i}{c_j}\bra{\Psi_j}H\ket{\Psi_i}\Big],0\Big)\ .&
\label{trt}
}
{\it The macrostate of the system is associated to the macroscopic time averages of $\bra{\Psi_{i(t)}}{\cal O}_n\ket{\Psi_{i(t)}}$.}
\vspace{0.2cm}
\hrule\end{changemargin}

\vspace{0.2cm}
The stochastic process defined by the transition rates \eqref{trt} allows only one way transitions between each pair of states, i.e.~if $\ket{\Psi_i}\to\ket{\Psi_j}$ has a non-vanishing rate then the reverse $\ket{\Psi_j}\to\ket{\Psi_i}$ has a vanishing rate, and vice-versa.

The transition rates \eqref{trt} ensure that the stochastic process is compatible with the Schr\"odin-ger equation and microscopic Born rule in the sense that if $p_{i|j}(t,t')$ is the integrated transition probability that state $\ket{\Psi_{j(t')}}$ evolves via the process to $\ket{\Psi_{i(t)}}$:
\EQ{
\begin{tikzpicture}
\node at (0,2) (a1) {$|c_i(t)|^2=\displaystyle\sum_jp_{i|j}(t,t')|c_j(t')|^2$};
\node at (-4,0) (a2) {$\ket{\Psi(t)}=\displaystyle\sum_ic_i(t)\ket{\Psi_i}$};
\node at (4,0) (a3) {$\ket{\Psi(t')}=\displaystyle\sum_ic_i(t')\ket{\Psi_i}$};
\node at (0,0.7) (a4) {\scriptsize $H\ket{\Psi}=i\hbar\dfrac d{dt}\ket{\Psi}$};
\draw[<-,very thick] (-1.6,0.15) -- (1.5,0.15);
\draw[<->] (-3.2,0.6) -- (-3.2,1.2) -- (-2.3,1.2) -- (-2.3,1.8);
\draw[<->] (4.6,0.6) -- (4.6,1.2) -- (1.8,1.2) -- (1.8,1.8);
\end{tikzpicture}
\label{ue1}
}
where the time dependence of $c_i(t)$ follows implicitly from underlying Schr\"odinger equation \eqref{hem}. It is interesting that, in our context, it is the distribution $|c_i(t)|^2$ that is fundamental while the stochastic process is slave to it, rather than the more conventional situation where the transition rates would given and one would have to solve \eqref{ue1} for the $|c_i(t)|^2$.
The result \eqref{ue1} means that, {\em if\/} the quantum microstates were distributed with probabilities $|c_i(t')|^2$ at time $t'$, then they will distributed with probabilities $|c_i(t)|^2$ at all later times $t>t'$. It might seem artificial to take this as an initial condition, after all the system really is, from the perspective of $\EA$, has expectation values $\bra{\Psi_{i(t')}}{\cal O}_n\ket{\Psi_{i(t')}}$ associated to a particular microstate at time $t'$. However,
if the system is in equilibrium and the process is ergodic, then the $|c_i|^2$ are approximately time independent and after a suitable time, $p_{i|j}(t,t')$ becomes effectively independent of $j$, i.e.~the process forgets its starting state. In this situation, $|c_i(t)|^2$ {\em does\/} become interpreted as the probability that the quantum microstate is $\ket{\Psi_{i(t)}}$ at time $t$ and the expectation values calculated from long time averages of the stochastic process are equal the underlying expectation value. 

On the other hand, it is important also that in other situations ergodicity of the process can be broken, or partially broken, in order that Newton's laws can emerge and the measurement problem can be solved.
There are three kinds of behaviour that are needed to successfully account for the classical limit of macro-systems:

\vspace{0.2cm}
\begin{changemargin}{0.3cm}{0.3cm}
\hrule
\vspace{0.2cm}{\it
\no{\bf 1. Deterministic:} this occurs when the matrix elements $\bra{\Psi_i}H\ket{\Psi_j}$ are vanishingly small  which can be expected when the quantum microstates are 
macroscopically distinct. In this case, over realistic macroscopic time scales the ergodicity of the stochastic process is broken. The possible stochastic trajectories then localize around the deterministic trajectories of classical mechanics. 

\vspace{0.2cm}
\no{\bf 2. Stochastic:} this occurs when the matrix elements $\bra{\Psi_i}H\ket{\Psi_j}$ are generic and the stochastic process is ergodic. This is exactly the situation needed to describe the thermal properties of macroscopic systems.

\vspace{0.2cm}
\no{\bf 3. Quantum measurement:} (a mixture of 1 and 2) during a measurement on a microscopic quantum system, the final state of the macroscopic measuring device splits up into subsets of quantum microstates $\EE_a\subset\MS$ for which the matrix elements $\bra{\Psi_i}H\ket{\Psi_j}$
with $\ket{\Psi_i}\in\EE_a$ and $\ket{\Psi_j}\in\EE_b$, for $a\neq b$, are vanishing small because the quantum microstates in different subsets are macroscopically distinct. The subsets $\EE_a$ become ergodically disjoint and therefore a definite measurement outcome occurs depending on which of the ergodic subsets the system lies in. 
In this context, the reduction of the state vector to the ergodic component corresponding to the measurement outcome that is actually realized, 
\EQ{
\sum_{\ket{\Psi_i}\in\MS}c_i\ket{\Psi_i}\longrightarrow\sum_{\ket{\Psi_j}\in\EE_b}c_j\ket{\Psi_j}\ ,
\label{bb2}
}
is just an efficient piece of book keeping to reflect the fact that over macroscopic time scales 
$T$ the integrated probabilities $p_{i|j}(t+T,t)$, for $\ket{\Psi_j}\in{\cal E}_b$ to $\ket{\Psi_i}\in{\cal E}_a$, $a\neq b$, are vanishingly small.
\vspace{0.2cm}
\hrule}\end{changemargin}

\section{Quantum Microstates}\label{s3}

In this section, we define the set of quantum microstates $\MS=\{\ket{\Psi_i}\}$ that are key for our approach. It is important to recognize that these states depend implicitly on both the underlying quantum state $\ket{\Psi}$ and the set of observables $\EA$. 

Following the classic work of Jaynes \cite{Jaynes2}, if we have incomplete knowledge of an underlying quantum state $\ket{\Psi}$ in the form of the expectation values of a set of observables $\bra{\Psi}{\cal O}_n\ket{\Psi}$, ${\cal O}_n\in\EA$, then the most unbiased description of the state is the density operator $\rho$ with
\EQ{
\bra{\Psi}{\cal O}_n\ket{\Psi}=\Tr\big(\rho{\cal O}_n\big)\ ,\qquad\forall\ {\cal O}_n\in\EA\ ,
\label{ll3}
}
which maximizes the von~Neumann entropy
\EQ{
S=-\Tr\big(\rho\log\rho\big)\ ,
}
subject to the constraints \eqref{ll3}.
In a sense, we can view $\rho$ as describing an ensemble of its eigenstates $\ket{\Psi_i}$,
\EQ{
\rho=\sum_i|c_i|^2\ket{\Psi_i}\bra{\Psi_i}\ ,
}
with the eigenvalues $|c_i|^2$ interpreted as probabilities. In the following, we will make this ensemble interpretation precise and will find that it emerges as the macrostate of a system in thermal equilibrium.

There are, of course, many density operators that can reproduce the expectation values of a set of observables $\EA$, but the maximal entropy requirement picks out a unique one.
We will make a refinement of this notion, by adding an additional constraint that we call the {\it microscopic Born rule}.
This leads us to propose the definition of the set of quantum microstates $\MS=\{\ket{\Psi_i}\}$:

\vspace{0.2cm}
\begin{changemargin}{0.3cm}{0.3cm}
\hrule
\vspace{0.2cm}
\no{\it
{\bf 1. Microscopic Born rule:} for each observable ${\cal O}_n\in\EA$
\EQ{
\bra{\Psi}{\cal O}_n\ket{\Psi}=\sum_i|c_i|^2\bra{\Psi_i}{\cal O}_n\ket{\Psi_i}\ ,\qquad c_i=\bra{\Psi_i}\Psi\rangle\ .
\label{sa1}
}

\no{\bf 2. Non-degeneracy:} for every pair of states $\ket{\Psi_k}$ and $\ket{\Psi_l}$, there exists at least one ${\cal O}_n\in\EA$ for which
\EQ{
\bra{\Psi_i}{\cal O}_n\ket{\Psi_j}\neq\mu_n\delta_{ij}\ ,\qquad i,j\in\{k,l\}\ .
\label{sa2}
}
If the contrary were true then we simply reduce the dimension of $\MS$ by replacing the pair $\ket{\Psi_k}$ and $\ket{\Psi_l}$ with the, suitably normalized, linear combination
\EQ{
c_k\ket{\Psi_k}+c_l\ket{\Psi_l}\ .
}
This condition, ensures that the quantum microstates are distinguishable with respect to the set of observables $\EA$ and it acts to limit the size of $\MS$.

\vspace{0.2cm}
\no{\bf 3. Maximal entropy:}  subject to the conditions above, the coarse grained density operator has maximal entropy
\EQ{
S=-\Tr\big(\rho\log\rho\big)=-\sum_i|c_i|^2\log|c_i|^2\ .
\label{kk3}
}
This condition, roughly speaking, acts to make the set $\MS$ as big as possible subject to the other constraints.
\vspace{0.2cm}
\hrule}\end{changemargin}

\vspace{0.2cm}
As an example, suppose that $\EA$ consists of a complete set of observables acting on a tensor product factor $\BH_A$ of a Hilbert space $\BH=\BH_A\otimes\BH_E$. In this case, discussed fully as Example 2 in appendix \ref{a3}, the quantum microstates $\ket{\Psi_i}$ are the states appearing in the Schmidt decomposition 
\EQ{
\ket{\Psi}=\sum_ic_i\ket{\Psi_i}\ ,\qquad\ket{\Psi_i}=\ket{\psi_i}\otimes\ket{\tilde\psi_i}\ .
\label{pf2}
}
These states form the basis for the reduced density operator formalism developed in \cite{Hollowood:2013cbr,Hollowood:2013xfa,Hollowood:2013bja}.\footnote{In those works, the states $\ket{\psi_i}$ of $\BH_A$ were called ontic states and $\ket{\tilde\psi_i}$ their mirrors. Here, given the more general context and the relation to statistical mechanics, we prefer the more general term {\it quantum microstate\/} for $\ket{\Psi_i}$ the states of $\BH_A\otimes\BH_E$.}

\subsection{Realistic sets of observables}\label{s3.1}

To start with, let us suppose that
$\EA$ to consists of a set of commuting operators. In this case, the quantum microstates are simply the mutual eigenstates that appear in the decomposition of $\ket{\Psi}$. It is clear that the microscopic Born rule \eqref{sa1} is satisfied.  More precisely, the non-degeneracy condition \eqref{sa2}, ensures that there is only one quantum microstate for each degenerate eigenspace $\BH_i\subset\BH$. The maximal entropy condition requires that each each degenerate eigenspace contributes separately.

In order to describe a macro-system, one expects that the set of observables $\EA$ contains coarse grained versions of the microscopic observables. Each such observable ${\cal O}$ has a finite working range $\Delta({\cal O})$ and a finite resolution $\delta_ {\cal O}$. We can define a set projection operators associated to the intervals $[\lambda_{i-1},\lambda_i]$ which cover the working range of ${\cal O}$ with $\lambda_i=\lambda_0+i\delta_{\cal O}$, $i=1,2,\ldots,N$, where $N\delta_{\cal O}=\Delta({\cal O})$, defined as
\EQ{
\Pi^{({\cal O})}_i=\sum_{\lambda_a\in[\lambda_{i-1},\lambda_i]}\ket{\psi_a}\bra{\psi_a}\ ,\qquad\Pi^{({\cal O})}_i\Pi^{({\cal O})}_j=\delta_{ij}\Pi^{({\cal O})}_i\ ,
}
where $\ket{\psi_a}$ and $\lambda_a$ are the eigenstates and eigenvalues of ${\cal O}$. For operators with continuous spectra, the sum is replaced by an integral. One can then define a class of coarse grained observables by taking linear combinations of these projection operators
$\sum_i\alpha_i\Pi^{({\cal O})}_i$, for example a coarse grained version of ${\cal O}$ can be written
\EQ{
{\cal O}_{\text{c.g}}=\sum_{i=1}^N\big(\lambda_0+(i-\tfrac12)\delta_{\cal O}\big)\Pi^{({\cal O})}_i\ .
}
These operators have a discrete spectrum and are generally highly degenerate. A quantum microstate is associated to each degenerate subspace $\BH_i$ of the form
\EQ{
\ket{\Psi_i}=\sum_{\lambda_a\in[\lambda_{i-1},\lambda_i]}f_{ia}\ket{\psi_a}\ .
}
for coefficients $f_{ia}$ determined by the state $\ket{\Psi}$.

Of course, we expect that the set $\EA$ will contain many observables ${\cal O}_n$ and then the issue becomes how to treat the situation where the microscopic observables do not commute. If one builds the coarse grained observables out of the projection operators defined above then these projection operators will not commute. As an example, consider the position and momentum of a particle with associated projection operators
\EQ{
\Pi^{(x)}_i=\int_{x_{i-1}}^{x_{i}}dx\,\ket{x}\bra{x}\ ,\qquad \Pi^{(x)}_i\Pi^{(x)}_j=\delta_{ij}\Pi^{(x)}_i\ ,
\label{xpf}
}
for $x_i=x_0+i\delta_x$, and 
\EQ{
\Pi^{(p)}_i=\int_{p_{i-1}}^{p_i}dp\,\ket{p}\bra{p}\ ,\qquad \Pi^{(p)}_i\Pi^{(p)}_j=\delta_{ij}\Pi^{(p)}_i\ ,
}
with $p_i=p_0+i\delta_p$. These projection operators do not commute even though in the realistic limit $\delta_x\delta_p\gg\hbar$, the lack of commutation is small. 

In our approach there is nothing to prevent us building coarse grained observables on these projection operators leading to a non-abelian set $\EA$, as we describe in appendix \ref{a3}. Although this creates no conceptual difficulties, it does create technical difficulties in finding explicit expression for the quantum microstates. However, we can use the freedom that we have in defining the observables to slightly deform the operators $\Pi^{(x)}_i$ and $\Pi^{(p)}_j$ to get new coarse grained projection operators which {\it do\/} commute. Intuitively one expects that this should be possible when the volume of the cell in phase space $\delta_x\delta_p\gg\hbar$. The idea of deforming the position and momentum operators so that they commute, but which are suitably close to the original operators, goes back to von~Neumann \cite{vNb,vN}. For example, Halliwell \cite{Hall} has made an explicit construction of operators which are effectively deformations of our phase space projectors which do commute and are still projectors and so would be suitable for our purposes.\footnote{It is important that the projectors of Halliwell are not exhaustive on phase space to avoid the consequences of the Balian-Low theorem. However, this is not a restriction for us because there is no requirement that $\EA$ is generated by an exhaustive set of projection operators. It is more natural for us that $\EA$ consists of operators that only cover a local region of phase space.}
If we now take $\EA$ to consist of the deformed operators the quantum microstates are just the simultaneous eigenstates. These eigenstates are very accurately---but not exactly---localised in the cell $[x_{i-1},x_i]\cap[p_{j-1},p_j]$ in phase space when $\delta_x\delta_p\gg\hbar$. It is reasonable to expect that working with a slightly different set of observables $\EA$ will not lead to any significant differences in the resulting analysis.\footnote{This kind of freedom is familiar in effective QFT. For example, one can define coarse grained observables by defining the theory on a lattice or one can work in the continuum and cut off Fourier modes at high frequency. The IR behaviour of the QFT is insensitive to these differences in regularization.}

The same kind of deformation can be generalized to any non-abelian set $\EA$ where the non-commutativity is small and so it is sufficient to assume that the set of observables $\EA$ needed to describe the classical limit of a quantum system are abelian.

\subsection{Local observables}\label{s3.2}

In general, we expect that $\EA$ contains coarse-grained observables that are localized in some 
compact region of space $A$. To reflect this, the Hilbert space has a tensor product decomposition $\BH=\BH_A\otimes\BH_E$. The set of accessible operators $\EA$ act locally on the $\BH_A$ factor, i.e.~are of the form ${\cal O}_n\otimes1_E$. If there were no coarse graining, so that $\EA$ were a complete set of observables on $\BH_A$, then this would correspond to Example 2 in appendix \ref{a3}. More realistically we expect some amount of coarse graining so that the set $\EA$ is not complete. In this more realistic situation, therefore, the set of observables $\EA$ has both a ultra-violet cut off, the scale of the coarse graining, and an infra red cut off, the scale of the compact region in space.

Taking $\EA$ to consist of a set of commuting observables, the quantum microstates will generally be entangled with states of the environment, i.e.~of the form
\EQ{
\ket{\Psi}=\sum_{i=1}^Nc_i\ket{\Psi_i}\ ,\qquad\ket{\Psi_i}=\sum_{a}\ket{\psi_{ia}}\otimes\ket{\tilde\psi_{ia}}\ ,
\label{pf3}
} 
where $\ket{\psi_{ia}}$ are the orthonormal eigenstates of the abelian set $\EA$ with $a=1,2,\ldots,\text{dim}\,\BH_i$.\footnote{For the continuous spectrum case the sum is replaced by an integral.} These states look superficially like the Schmidt states \eqref{pf2}, however, there is no requirement here that the states $\ket{\tilde\psi_{ia}}$ are orthogonal. The only constraint on them is the normalization condition
\EQ{
\sum_{a}\bra{\tilde\psi_{ia}}\tilde\psi_{ia}\rangle=1\ .
}

\section{The Stochastic Process and Ergodicity}\label{s4}

In this section we discuss in more detail the stochastic process defined by the transition rates \eqref{trt}. The first point, 
amplifying what we said earlier in section \ref{s2}, the process is consistent with the Schr\"odinger equation in the sense that if we temporarily were to interpret $|c_i|^2$ as the probability that the system from the point of view of ${\EuScript A}$ is in the quantum microstate $\ket{\Psi_i}$ at time $t$, then the stochastic process preserves this interpretation at later time, as in \eqref{ue1}. So viewing the quantum microstates as an ensemble with probabilities $|c_i|^2$ then the
ensemble average over the quantum microstates evolving according to the stochastic process 
equals the ordinary expectation value for any observable in $\EA$.

From the Schr\"odinger equation, we find
\EQ{
\frac{d|c_i|^2}{dt}=-\frac2\hbar\sum_{j\neq i}\IM\Big[c_ic_j^*\bra{\Psi_j}\Big(H-i\hbar\frac d{dt}\Big)\ket{\Psi_i}\Big]\ .
\label{eqj}
}
In realistic cases, the observables $\EA$ are time independent and it follows that
\EQ{
\bra{\Psi_j}\frac d{dt}\ket{\Psi_i}=0\ ,\qquad i\neq j\ .
}
The generalization to the time dependent case is described in appendix \ref{a2}.
Equation \eqref{eqj} can be written in the form of a master equation of a stochastic process
\EQ{
\frac{d|c_i|^2}{dt}=\sum_{j\neq i}\Big[T_{ij}\,|c_j|^2-T_{ji}\,|c_i|^2\Big]\ ,
\label{ge2}
}
where $T_{ij}$ is the transition rate for $\ket{\Psi_j}\to\ket{\Psi_i}$.

Our task is to solve \eqref{ge2} given the known time dependence of $c_i(t)$ that follow from the Schr\"odinger equation in \eqref{eqj}. Note that in the theory of stochastic processes, one would usually be presented with the problem the other way around, i.e.~given the transition rates $T_{ij}$ one would have to solve for the instantaneous probabilities $|c_i(t)|^2$. There is a very natural solution for the transition rates of the form\footnote{It might be thought unnecessary to define the stochastic process as a continuous process in time and indeed in earlier work \cite{Hollowood:2013cbr,Hollowood:2013xfa} we 
defined the process with some temporal cut off chosen to be much smaller than any characteristic time scales in the problem. This was done to avoid a problem with ``crossovers" discussed in detail in \cite{Hollowood:2013cbr,Hollowood:2013xfa}. But the problem with crossovers only arises from taking an unrealistic choice for the observables as in Example 2 in appendix \ref{a3}.
Realistic choices as described in section \ref{s3.1} do not suffer from this problem. So in the end whether to take a continuous or discrete version of the stochastic process is simply a matter of computational expedience.}
\EQ{
T_{ij}=\text{max}\Big(-\frac2\hbar\IM\Big[\frac{c_i}{c_j}\bra{\Psi_j}H\ket{\Psi_i}\Big],0\Big)\ ,
\label{trt10}
}
which is the expression quoted in \eqref{trt}. Note that if $T_{ij}$ is non-vanishing, then $T_{ji}=0$ and vice-versa. The form makes clear that it is the mis-alignment between the energy eigenstate basis and the quantum microstate basis that drives the transitions.

The transition probabilities \eqref{trt10}  are actually not the most general set of probabilities that are consistent with \eqref{eqj}. The freedom involves $T_{ij}\to T_{ij}+\Omega_{ij}/|c_j|^2$ for any symmetric $\Omega_{ij}>0$. However, it is our hypothesis that $\Omega_{ij}=0$ because $T_{ij}$ is then not only the simplest and most natural choice depending as it does on just the matrix elements between the initial and final quantum microstates $\bra{\Psi_j}H\ket{\Psi_i}$ but also the choice has some important implications for the ergodicity of the stochastic process that will be important for obtaining a sensible classical limit.\footnote{The transition rates \eqref{trt10} are identical to  
written down by Bell in his theory of beables \cite{Bell:2004suqm} while the generalization with time dependent $\ket{\Psi_i}$ described  in appendix \ref{a2} was considered by Bacciagaluppi and Dickson \cite{BacciagaluppiDickson:1999dmi}.
In Bell's case, the $\ket{\Psi_i}$ were taken to be eigenstates of some set of  preferred observables, the {\it beables\/}. In our approach, the set of observables is not fixed once and for all in some global way, rather it is dictated locally by the way the quantum system interacts macroscopically with it surroundings.}

\subsection{Ergodicity and the classical limit}\label{s4.1}

The key feature of the stochastic process is that it preserves the equality of the ensemble average with respect to the coarse grained density operator $\rho$ with the expectation value \eqref{kww}. The question is how is the quantum expectation value $\bra{\Psi}{\cal O}\ket{\Psi}$, related to the time sequence of expectation values 
$\bra{\Psi_{i(t)}}{\cal O}_n\ket{\Psi_{i(t)}}$ of the stochastic process?
The situation here is very similar to the relation between ensemble averages and long time averages in classical SM. 
The issue rests on the ergodicity of the stochastic process. In the ergodic situation ensemble averages capture long time averages. In classical SM, it is generally very difficult to establish ergodicity because classical mechanics is deterministic and classical motion corresponds to a trajectory in phase phase. It is then a delicate matter to argue that the trajectory samples phase space in an ergodic way. The present quantum situation is much simpler, partly because the quantum microstates are discrete, and, partly because they follow stochastic dynamics. Moreover, the dynamics is a conceptually simple continuous time Markov chain for which ergodicity is---conceptually at least---straightforward to establish. 

The issue of ergodicity, when it holds and when it is broken, is the key to understanding how a classical world can emerge from quantum mechanics in our proposal.
When the system is in thermal equilibrium, the expectation values $\bra{\Psi}{\cal O}_n\ket{\Psi}$ are approximately constant. We then argue in section \ref{s7} that the probabilities $|c_i(t)|^2$ are also approximately constant, a fact that actually follows from von~Neumann's quantum ergodic theorem \cite{vN}. The stochastic process 
will effectively be ergodic over a long time $T$ if the 
integrated transition probabilities $p_{i|j}(t+T,t)$ become independent of $j$ so that memory of the initial state is lost. It follows from \eqref{ue1} that $|c_i|^2$ is precisely the probability that the system has expectation values $\bra{\Psi_{i}}{\cal O}_n\ket{\Psi_{i}}$. In particular, from a macroscopic point of view which involves a long time average compared with microscopic time scales, the long time average is captured by the ensemble average over the quantum microstates $\MS$ with probabilities $|c_i|^2$: 
\begin{flalign}
&\text{\bf Equilibrium:}\qquad 
\frac1T\int_0^T dt\,\bra{\Psi_{i(t)}}{\cal O}_n\ket{\Psi_{i(t)}}
\approx\sum_i|c_i|^2\bra{\Psi_i}{\cal O}_n\ket{\Psi_i}=\bra{\Psi}{\cal O}_n\ket{\Psi}\ ,&
\label{tt6}
\end{flalign}
for large enough $T$. In the above, the last equality is the microscopic Born rule \eqref{sa1}. We will investigate ergodicity in the context of SM more fully in section \ref{s7}.

However, we do not expect that the stochastic process of a system will always be ergodic; indeed if it were so then the classical limit would not involve any deterministic dynamics. So the issue of ergodicity breaking is key to understanding how macroscopic systems can obey Newton's laws.
Ergodicity breaking occurs when pairs of quantum microstates $\ket{\Psi_i}$ and $\ket{\Psi_j}$ are macroscopically distinct because it would then follow that the matrix elements $\bra{\Psi_j}H\ket{\Psi_i}$ are vanishingly small and so the transition rate \eqref{trt} between the two states and the integrated probability $p_{i|j}(t,t')$ would be hugely suppressed and ergodicity over macroscopic time scales would inevitably be broken. For any given initial quantum microstate $\ket{\Psi_i}$, the stochastic process would then effectively only explore a more limited ergodic subspace of $\MS$. This subspace consists of quantum microstates that are tightly clustered around a classical trajectory, as we illustrate in section \ref{s6}.

A macroscopic object will typically have elements of ergodic and non-ergodic behaviour: the position of its moving parts in space will involve a breaking of ergodicity whereas, at the same time, it will typically be in thermal equilibrium, behaviour that involves an ergodic component corresponding to the appropriate thermal ensemble.

\subsection{Locality}\label{s5}

The issue of locality in quantum mechanics is full of misunderstandings.\footnote{We refer to \cite{Englert} for an insightful discussion of these kinds of misunderstandings. In particular, quantum mechanics famously violates Bell's theorem but is completely consistent with experiments. So the theorem holds for theories that are not realized in nature. It is hard to see that it has any implications for quantum mechanics itself and, in particular, it most certainly does not imply that quantum mechanics is non-local. All ``delayed choice" type experiments designed, apparently, to highlight the spooky, non-local nature of quantum mechanics simply reveal the correlations between entangled states: there is no non-locality.} It cannot be over emphasised that quantum mechanics is a perfectly local theory. This is particular clear in QFT where locality and causality are built into the fabric of the formalism: interaction terms in the Hamiltonian are spatially local; and causality is manifested in the fact that space-like separated operators commute.

The worry might be that the stochastic process we have introduced some non-locality. 
We need to show that the stochastic dynamics associated to a set of observables $\EA=\{{\cal O}_n\}$ that act on spatially localized degrees of freedom is only driven by interactions that are local to these degrees of freedom. 
In the case of a spatially localized subsystem $A$, we can expect the Hilbert space to factorize $\BH=\BH_A\otimes\BH_E$ with the observables acting as ${\cal O}_n\otimes 1_E$ and a Hamiltonian with the form
\EQ{
H=H_A\otimes 1_E+1_A\otimes H_E+H_\text{int}\ .
}

We will consider the physically realistic situation where $\EA$ consists of a set of commuting operators acting on $\BH_A$ factor described in section \ref{s3.2}. In this case, the quantum microstates on the total system take the form \eqref{pf3}, where the states $\ket{\psi_{ia}}$ are the simultaneous eigenstates of the commuting set $\EA$. The key point is that the set of states $\{\ket{\psi_{ia}}\}$ are orthogonal while the set $\{\ket{\tilde\psi_{ia}}\}$ are not. Therefore, 
\EQ{
\bra{\Psi_i}1_A\otimes H_E\ket{\Psi_j}=0\ ,\qquad i\neq j\ ,
\label{pp2}
}
on account of the fact that $\bra{\psi_{ia}}\psi_{jb}\rangle=\delta_{ij}\delta_{ab}$;
hence, the transition rates take the form
\EQ{
T_{ij}=\text{max}\Big(-\frac2\hbar\IM\Big[\frac{c_i}{c_j}
\bra{\Psi_j}\Big\{H_A\otimes 1_E+H_\text{int}\Big\}\ket{\Psi_i}\Big],0\Big)\ .
\label{trt3}
}
It follows that the stochastic process depends on the internal dynamics of the subspace $A$, via $H_A$, as well as the local  interactions of $A$ with the environment $E$, via $H_\text{int}$. Importantly, though, the dynamics of the environment involving $H_E$ are irrelevant. This proves that the stochastic dynamics is local to the degrees of freedom on which the observables $\EA$ act.

\subsection{Patching together a classical world}

The discussion above has implications for the case when the set of operators ${\EuScript A}\cup{\EuScript B}$ where the subsets consists of local observables acting on spatially separated degrees of freedom. The question is how the perspectives of the system provided individually by ${\EuScript A}$ and ${\EuScript B}$ are related to the global perspective provided by ${\EuScript A}\cup{\EuScript B}$. If the macroscopic world is to be classical then it must be the different views can be integrated together, at least to high precision: a classical world is, after all, an approximation.

The first point is that the quantum microstates of ${\EuScript A}\cup{\EuScript B}$ can naturally be given a pair of labels $\ket{\Psi_{i_1i_2}}$ to reflect the decomposition of ${\EuScript A}\cup{\EuScript B}$. So the label $i$ is split into the pair $i_1,i_2$ which labels the eigenstates of the sets ${\EuScript A}$ and ${\EuScript B}$, respectively. The quantum microstates of the individual localized sets ${\EuScript A}$ or ${\EuScript B}$ are given by the partial sums
\EQ{
c_{i_1}\ket{\Psi_{i_1}}=\sum_{i_2}c_{i_1i_2}\ket{\Psi_{i_1i_2}}\ ,\qquad
c_{i_2}\ket{\Psi_{i_2}}=\sum_{i_1}c_{i_1i_2}\ket{\Psi_{i_1i_2}}\ ,
\label{pn1}
}
with
\EQ{
|c_{i_1}|^2=\sum_{i_2}|c_{i_1i_2}|^2\ ,\qquad|c_{i_2}|^2=\sum_{i_1}|c_{i_1i_2}|^2\ .
}
Notice that the partial sums in \eqref{pn1} are forced by the non-degeneracy condition \eqref{sa2} because
\EQ{
\bra{\Psi_{i_1i_2}}{\cal O}_n\otimes 1_{\EuScript B}\ket{\Psi_{i_1j_2}}=\mu_n\delta_{i_2j_2}\ ,\qquad\forall{\cal O}_n\in\EA\ .
}
When $c_{i_1i_2}\neq c_{i_1}c_{i_2}$ the state has entanglement between the two spatially separated regions. Entanglement could occur if the two subsystems were the two measuring devices in an EPR type experiment as we describe in section \ref{s9}. In this section we will focus on the non-entangled situation.

The stochastic process of the total set ${\EuScript A}\cup{\EuScript B}$ is just a refinement of the individual stochastic processes of the two subsets with
\EQ{
T^{({\EuScript A})}_{i_1j_1}=\sum_{i_2,j_2}T^{({\EuScript A}\cup{\EuScript B})}_{i_1i_2,j_1j_2}\Big|\frac{c_{j_1j_2}}{c_{j_1}}\Big|^2\ ,\qquad
T^{({\EuScript B})}_{i_2j_2}=\sum_{i_1,j_1}T^{({\EuScript A}\cup{\EuScript B})}_{i_1i_2,j_1j_2}\Big|\frac{c_{j_1j_2}}{c_{j_2}}\Big|^2\ .
}

In order that the two stochastic processes associated to ${\EuScript A}$ and ${\EuScript B}$ are genuinely independent requires that:

\begin{enumerate}
\item The only non-vanishing transitions of the global system are
$\ket{\Psi_{j_1i_2}}\to\ket{\Psi_{i_1i_2}}$ and $\ket{\Psi_{i_1j_2}}\to\ket{\Psi_{i_1i_2}}$.
\item The rate for $\ket{\Psi_{j_1i_2}}\to\ket{\Psi_{i_1i_2}}$ is independent of $i_2$ and that for $\ket{\Psi_{i_1j_2}}\to\ket{\Psi_{i_1i_2}}$ is independent of $i_1$.\footnote{This is violated in the entangled situation discussed in section \ref{s9} due to the correlations between the states.}
\end{enumerate}

\no These conditions together imply that the only non-vanishing transition rates of the total system are
\EQ{
T^{({\EuScript A}\cup{\EuScript B})}_{i_1i_2,j_1i_2}=T^{({\EuScript A})}_{i_1j_1}\ ,\qquad
T^{({\EuScript A}\cup{\EuScript B})}_{i_1i_2,i_1j_2}=T^{({\EuScript B})}_{i_2j_2}\ .
}
Recalling the formula for the transition rates \eqref{trt10}, the first condition above follows from the fact that realistic Hamiltonians only couple spatially local degrees of freedom and so cannot drive transitions at two spatially separated regions simultaneously; hence
\EQ{
\bra{\Psi_{i_1i_2}}H\ket{\Psi_{j_1j_2}}=0\qquad\text{if}\quad i_1\neq j_1\quad \text{and}\quad i_2\neq j_2\ .
}
The second condition above is more exacting and is obviously violated in the case where the systems are entangled due to the correlations. In the non-entangled case, let us consider the more detailed form of the quantum microstates. Let us take the Hilbert space to have the factorized form $\BH=\BH_1\otimes\BH_E\otimes\BH_2$ to reflect the spatial locality of the observables with elements of ${\EuScript A}$ acting as ${\cal O}\otimes 1_E\otimes 1_2$ and elements of ${\EuScript B}$ as $1_1\otimes 1_E\otimes{\cal O}$. The subsystem $\BH_E$ is the environment. We can write the quantum microstates of ${\EuScript A}\cup{\EuScript B}$, generalizing \eqref{pf3}, more concretely as
\EQ{
\ket{\Psi_{i_1i_2}}=\sum_{a_1a_2}\ket{\psi_{i_1a_1}}\otimes\ket{\tilde\psi_{i_1a_1,i_2a_2}}\otimes
\ket{\phi_{i_2a_2}}\ ,
}
which shows that the subsystems are generally entangled with the environment. Note that the sets $\{\ket{\psi_{i_1a_1}}\}$ and $\{\ket{\phi_{i_2a_2}}\}$ are orthonormal but the set $\{\ket{\tilde\psi_{i_1a_1,i_2a_2}}\}$ is not. So even though $H$ cannot couple degrees of freedom in $\BH_1$ and $\BH_2$ directly, the question is whether it can couple degrees of freedom in $\BH_1$
and those in the environment that are entangled with those in $\BH_2$, or vice-versa? 

Any such coupling must be very small because, by locality, degrees of freedom in $\BH_1$ can only couple locally with the environment and
the latter is a much bigger system than the subsystems $\BH_1$ and $\BH_2$ and it is unlikely 
the degrees of freedom of the environment that are entangled with $\BH_2$ are also spatially local to $\BH_1$. So, for realistic systems, to a high degree of accuracy, the second condition above is likely fulfilled and the stochastic process associated to ${\EuScript A}\cup{\EuScript B}$ consists of two independent processes ${\EuScript A}$ and ${\EuScript B}$ and the perspectives of ${\EuScript A}$ and ${\EuScript B}$ can be integrated into the global view ${\EuScript A}\cup{\EuScript B}$.

We will discuss more aspects of situation with two sets of spatially separated observables with entanglement in section \ref{s9}.

\section{A Particle}\label{s6}

In this section, we develop our formalism for the case of a particle moving in one dimension, although the generalization to more dimensions is then obvious. In this crude model, we will simply ignore the internal structure of the particle and take it to be a quantum particle with a wave function $\Psi(x)$. We will also simplify the situation even further by taking $\EA$ is consist
of a suitably coarse grained definition of the position operator only. The momentum will be added in later.

The observables consist of the set ${\EuScript A}=\{\Pi^{(x)}_i\}$ defined in \eqref{xpf}. In this case, the observables are mutually commuting and so the quantum microstates are simultaneous eigenstates. More precisely, the 
eigenstates have the degenerate eigenspaces that are are spanned by the 
eigenstates $\ket{x}$ in each interval $x\in[x_{i-1},x_i]$, with
\EQ{
\Pi^{(x)}_i\ket{x}=\begin{cases}\ket{x} &   x\in[x_{i-1},x_i]\ ,\\ 0 &\text{otherwise}\ .\end{cases}
}
The quantum microstates follow from the decomposition of $\ket{\Psi}$:
\EQ{
\ket{\Psi}=\sum_ic_i\ket{\Psi_i}\ ,\qquad
\ket{\Psi_i}=\frac1{c_i}\int_{x_{i-1}}^{x_i}dx\,\Psi(x)\ket{x}\ ,
}
with
\EQ{
|c_i|^2=\int_{x_{i-1}}^{x_i}dx\,\big|\Psi(x)\big|^2\ .
}
The quantum microstates in this case, are spatially localized an the interval $[x_{i-1},x_i]$ and represent states that have a good classical interpretation on scales $\gg\delta_x$.

\subsection{The stochastic process}

The transition rates are given in \eqref{trt10} and they have been calculated by Sudbery in \cite{SudJPA} whose calculation we repeat. We note that
\EQ{
c_j\Psi_j(x)=\big(\theta(x-x_{j-1})-\theta(x-x_j)\big)\Psi(x)
}
and therefore the matrix elements $\bra{\Psi_j}H\ket{\Psi_i}$ are only non-vanishing if $i=j\pm1$. Taking $i=j+1$, we have
\EQ{
\IM\Big[\frac{c_{j+1}}{c_{j}}\bra{\Psi_j}H\ket{\Psi_{j+1}}\Big]&=
-\frac{\hbar^2}{M|c_j|^2}\int dx\,\theta(x-x_j)\delta(x-x_j)\IM\Big[\Psi^*\frac{\partial\Psi}{\partial x}\Big]\ ,
}
with a similar expression for $i=j-1$. Evaluating the integral\footnote{Using $\int dx\,\theta(x-x_j)\delta(x-x_j)f(x)=\frac12f(x_j)$.} gives
\EQ{
\IM\Big[\frac{c_i}{c_j}\bra{\Psi_j}H\ket{\Psi_i}\Big]=\frac{\hbar^2}{2M|c_j|^2}\IM\Big(\Psi^*\frac{\partial\Psi}{\partial x}\Big|_{x=x_{j-1}}
\delta_{i,j-1}-\Psi^*\frac{\partial\Psi}{\partial x}\Big|_{x=x_j}\delta_{i,j+1}\Big)\ .
\label{pg2}
}
So the non-vanishing transitions rates involve $\ket{\Psi_j}\to\ket{\Psi_{j\pm1}}$ with \cite{SudJPA}
\EQ{
T_{j\pm1,j}=\text{max}\left(\pm\frac\hbar{M|c_j|^2}\IM\Big[\Psi^*\frac{\partial\Psi}{\partial x}\Big|_{x=x_j,x_{j-1}}\Big],0\right)\ .
\label{pg3}
}
The expression in the brackets here is, of course, just the Schr\"odinger probability current evaluated at the interface between 
the $j^\text{th}$ cell and one of adjacent cells divided by the probability to be $j^\text{th}$ cell. The transitions only go in the direction of the current.

We can now investigate this process in certain interesting limits. 
An important issue to consider is the relative magnitude of the width of the wave function of the particle $\Delta$ and the resolution scale $\delta_x$. For a macroscopic object, we expect that the wave function will be very narrow compared with the coarse graining scale $\delta_x\gg\Delta$. We will consider this case first. However, it is also interesting to consider situations where the wave function is smooth on the resolution scale $\delta_x\ll\Delta$. 

\subsection{Narrow wave function $\delta_x\gg\Delta$}\label{s6.1}

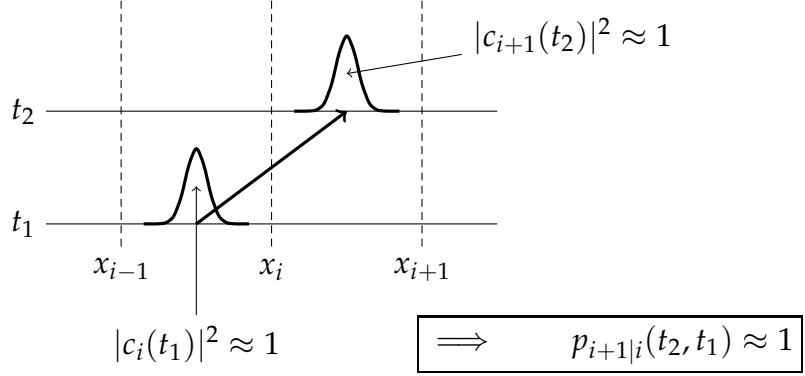
\begin{figure}[ht]
\begin{center}
\begin{tikzpicture}[scale=1]
\draw[-] (0,0) -- (6,0);
\draw[-] (0,1.5) -- (6,1.5);
\draw[densely dashed] (1,-0.3) -- (1,3);
\draw[densely dashed] (3,-0.3) -- (3,3);
\draw[densely dashed] (5,-0.3) -- (5,3);
\begin{scope}[xscale=0.1,yscale=1,xshift=20cm,yshift=0cm]
\draw[-,very thick] plot[smooth] coordinates {(-7, 0) (-6, 0) (-5., 0.00193045) (-4., 
  0.0183156) (-3., 0.105399) (-2., 0.367879) (-1., 0.778801) (0., 
  1.) (1., 0.778801) (2., 0.367879) (3., 0.105399) (4., 
  0.0183156) (5., 0.00193045) (6., 0) (7., 0)};
\end{scope}
\begin{scope}[xscale=0.1,yscale=1,xshift=40cm,yshift=1.5cm]
\draw[-,very thick] plot[smooth] coordinates {(-7, 0) (-6, 0) (-5., 0.00193045) (-4., 
  0.0183156) (-3., 0.105399) (-2., 0.367879) (-1., 0.778801) (0., 
  1.) (1., 0.778801) (2., 0.367879) (3., 0.105399) (4., 
  0.0183156) (5., 0.00193045) (6., 0) (7., 0)};
\end{scope}
\node at (-0.3,0) (a1) {$t_1$};
\node at (-0.3,1.5) (a1) {$t_2$};
\node at (1,-0.6) (a1) {$x_{i-1}$};
\node at (3,-0.6) (a1) {$x_i$};
\node at (5,-0.6) (a1) {$x_{i+1}$};
\node at (7.5,-1.6) (a1) {$\boxed{\implies\qquad p_{i+1|i}(t_2,t_1)\approx1}$};
\node at (2,-1.6) (a2) {$|c_i(t_1)|^2\approx1$};
\node at (7,2.5) (a3) {$|c_{i+1}(t_2)|^2\approx1$};
\draw[->] (a2) -- (2,0.5);
\draw[->] (a3) -- (4,2);
\draw[very thick,->] (2,0) -- (4,1.5);
\end{tikzpicture}
\end{center}
\caption{\small When the wave function is narrow compared with the resolution scale, the quantum microstate just follows the peak of $|\Psi(x)|^2$ and the stochastic process effectively leads to deterministic dynamics.}
\label{f1}
\end{figure}

This limit, which is relevant to a macroscopic particle,  is very simple because if at some time $t_1$
the centre of the wave packet lies within a region $[x_{i-1},x_i]$, the quantum microstate $\ket{\Psi_i}$ 
has a probability that is $|c_i(t_1)|^2\approx1$. Essentially in this case, the underlying state $\ket{\Psi}\approx\ket{\Psi_i}$ is the only quantum microstate that has any appreciable probability.
The centre of wave packet will, on account of Ehrenfest's theorem, follow a classical trajectory $x(t)$ and so as the wave packet moves across into a neighbouring region $[x_i,x_{i+1}]$, at, say time $t_2$, the quantum microstate changes to $\ket{\Psi_{i+1}}$ with a probability close to unity. It is clear, therefore, that with a probability close to unity, the resulting sequence of quantum microstates just describes a coarse graining of the classical trajectory; that is, at time $t$ the microstate is $\ket{\Psi_{i(t)}}$ where $i(t)$ is determined by the condition $x(t)\in[x_{i(t)-1},x_{i(t)}]$. This is illustrated in figure \ref{f1}.

\subsection{Smooth wave function $\delta_x\ll\Delta$}\label{s6.2}

The other limit, occurs when the width of the wave function
is much greater than the resolution scale $\delta_x\ll\Delta$. In this case, the resulting dynamics is fundamentally stochastic. However, even in this case, if  the wave function has the typical WKB form
\EQ{
\Psi(x)=R(x)e^{iS(x)/\hbar}\ ,
\label{ju3}
}
then a classical trajectory emerges. 
In the semi-classical limit $S(x)$, which is a smooth function on macroscopic scales, is identified with the classical action of the particle and the velocity of a classical trajectory is simply
\EQ{
v=\frac1M\frac{\partial S}{\partial x}\ .
}
We are assuming that the amplitude
$R(x)$ is a smooth function on the scale $\delta_x$. In this case,
the probabilities associated to the quantum microstates $\ket{\Psi_i}$ are $|c_j|^2\approx R(x_j)^2\delta_x$.

The transition rates \eqref{pg3} are exact but now we can use the form \eqref{ju3} for the wave function and the approximation $|c_j|^2\approx R(x_j)^2\delta_x$. This yields the transition probabilities
\EQ{
&T_{j-1,j}\approx\frac1{M\delta_x}\big|S'(x_j)\big|\ , \quad T_{j+1,j}=0\qquad \text{if}\quad S'(x_j)<0\ ,\\
&T_{j-1,j}=0\ ,\quad T_{j+1,j}\approx\frac1{M\delta_x}S'(x_j) \qquad~~~ \text{if}\quad S'(x_i)>0\ .
}
Given these transitions rates, it is clear that the stochastic process is quite simple: the particle can hop from the interval $[x_{j-1},x_j]$ to a neighbouring one, depending on the sign of $S'(x_j)$. Notice that $\hbar$ has completely dropped out of the expression for $T_{ij}$. 
In fact this stochastic process of the type we have derived is identical to the one considered by Vink \cite{Vink:1990fm} whose starting point and general approach is rather different,\footnote{In Vink's approach, the set up is rather different with the wave function being coarse grained on a spatial lattice whose spacing is taken to zero at the end. In our approach, our coarse graining scale is the spatial resolution scale which is kept finite.} but whose analysis we can still borrow.

The question is whether the resulting stochastic process gives a behaviour for the position that is recognizably classical. This does not seem at all obvious because classically there should just be a single trajectory. What we will argue is that the stochastic trajectories are clustered around the classical trajectory. 

In order to investigate the process
we introduce a small temporal cut off $\delta t$ and then investigate the process starting at, say $x=x_i$ over a time interval $T$ which involves a large number $N=T/\delta t$ of time steps, but is not too large, so that $S'(x_j)=vM$, $|i-j|<N$, is approximately constant and let us suppose $v>0$. For each time step, we have must have
\EQ{
p\equiv p_{i+1,i}=\frac{v}{\delta_x}\delta t\ll1\ ,
\label{pl6}
}
which can be achieved by having a small enough $\delta t$.
After the $N$ steps, the position $x_j$ with $j=k+i$ has a binomial probability distribution
\EQ{
\MAT{N\\ k}p^k(1-p)^{N-k}
}
whose average position is
\EQ{
\langle x\rangle=x_i+vT\ ,
}
which shows that $v$ is, indeed, interpreted as the velocity of the average of the distribution. 
The variance of the distribution is
\EQ{
\langle x^2\rangle-\langle x\rangle^2=vT\delta_x\big(1-\frac{v}{\delta_x}\delta t\big)\approx X\delta_x\ ,
}
where $X=vT$ is the distance moved in our small interval of time and we have used \eqref{pl6}. So for sub-macroscopic resolution $\delta_x$ the variance is small.

What we have shown here is that even when the wave function is wide compared with the resolution a classical trajectory can emerge. This suggests that there should be a way to relate the situation with a fixed wave function with different resolutions scales by changing the Heisenberg cut.
 
\subsection{Changing the Heisenberg cut}
 
One could view the results of sections \ref{s6.1} and \ref{s6.2} as pertaining to a given wave packet with fixed width but with a coarse and finer resolution scale $\delta_x$, respectively. 
A coarser view of the dynamics appears more deterministic than the finer view. The question then is, more generally, how does the stochastic dynamics behave as one
varies the Heisenberg cut. In the present example, there is a simple way to do 
this which involves a kind of blocking transformation that doubles the Heisenberg cut $\delta_x'=2\delta_x$. The new coarser description involves projection operators
\EQ{
\Pi^{(x)\prime}_i=\Pi^{(x)}_{2i-1}+\Pi^{(x)}_{2i}\ .
}
It follows that the new quantum microstates are simply
\EQ{
c'_i\ket{\Psi'_{i}}=c_{2i-1}\ket{\Psi_{2i-1}}+c_{2i}\ket{\Psi_{2i}}\ .
}
where $|c'_i|^2=|c_{2i-1}|^2+|c_{2i}|^2$.
The blocking transformation has a simple effect on the stochastic dynamics; the transition rates of the new process are related to the old one via
\EQ{
T_{ij}'=\frac1{|c'_j|^2}\big(T_{2i-1,2j-1}|c_{2j-1}|^2+T_{2i-1,2j}|c_{2j}|^2+T_{2i,2j-1}|c_{2j-1}|^2+T_{2i,2j}|c_{2j}|^2\big)\ ,
}

Performing this kind of blocking transformation allows us to the relate the two distinct regimes in sections \ref{s6.1} and \ref{s6.2}.
\begin{figure}[ht]
\begin{center}
\begin{tikzpicture}[scale=0.6,fill=black!20]
\begin{scope}[xshift=-3.5cm]
\draw[step=.4cm,gray,very thin] (-2.3,-2.3) grid (2.3,2.3);
\begin{scope}[xshift=-2.3cm,yshift=-2.0cm]
\draw[very thin,fill] (0,0) -- (0.3,0) -- (0.3,0.4) -- (0,0.4) -- (0,0);
\end{scope}
\begin{scope}[xshift=-2.0cm,yshift=-2.0cm]
\draw[very thin,fill] (0,0) -- (0.4,0) -- (0.4,0.4) -- (0,0.4) -- (0,0);
\end{scope}
\begin{scope}[xshift=-1.6cm,yshift=-2.0cm]
\draw[very thin,fill] (0,0) -- (0.4,0) -- (0.4,0.4) -- (0,0.4) -- (0,0);
\end{scope}
\begin{scope}[xshift=-1.2cm,yshift=-2cm]
\draw[very thin,fill] (0,0) -- (0.4,0) -- (0.4,0.4) -- (0,0.4) -- (0,0);
\end{scope}
\begin{scope}[xshift=-1.2cm,yshift=-1.6cm]
\draw[very thin,fill] (0,0) -- (0.4,0) -- (0.4,0.4) -- (0,0.4) -- (0,0);
\end{scope}
\begin{scope}[xshift=-0.8cm,yshift=-1.6cm]
\draw[very thin,fill] (0,0) -- (0.4,0) -- (0.4,0.4) -- (0,0.4) -- (0,0);
\end{scope}
\begin{scope}[xshift=-0.8cm,yshift=-1.2cm]
\draw[very thin,fill] (0,0) -- (0.4,0) -- (0.4,0.4) -- (0,0.4) -- (0,0);
\end{scope}
\begin{scope}[xshift=-0.8cm,yshift=-0.8cm]
\draw[very thin,fill] (0,0) -- (0.4,0) -- (0.4,0.4) -- (0,0.4) -- (0,0);
\end{scope}
\begin{scope}[xshift=-0.4cm,yshift=-0.8cm]
\draw[very thin,fill] (0,0) -- (0.4,0) -- (0.4,0.4) -- (0,0.4) -- (0,0);
\end{scope}
\begin{scope}[xshift=-0.4cm,yshift=-0.4cm]
\draw[very thin,fill] (0,0) -- (0.4,0) -- (0.4,0.4) -- (0,0.4) -- (0,0);
\end{scope}
\begin{scope}[xshift=-0cm,yshift=-0.4cm]
\draw[very thin,fill] (0,0) -- (0.4,0) -- (0.4,0.4) -- (0,0.4) -- (0,0);
\end{scope}
\begin{scope}[xshift=-0cm,yshift=-0cm]
\draw[very thin,fill] (0,0) -- (0.4,0) -- (0.4,0.4) -- (0,0.4) -- (0,0);
\end{scope}
\begin{scope}[xshift=0cm,yshift=0.4cm]
\draw[very thin,fill] (0,0) -- (0.4,0) -- (0.4,0.4) -- (0,0.4) -- (0,0);
\end{scope}
\begin{scope}[xshift=0.4cm,yshift=0.4cm]
\draw[very thin,fill] (0,0) -- (0.4,0) -- (0.4,0.4) -- (0,0.4) -- (0,0);
\end{scope}
\begin{scope}[xshift=0.4cm,yshift=0.8cm]
\draw[very thin,fill] (0,0) -- (0.4,0) -- (0.4,0.4) -- (0,0.4) -- (0,0);
\end{scope}
\begin{scope}[xshift=0.4cm,yshift=1.2cm]
\draw[very thin,fill] (0,0) -- (0.4,0) -- (0.4,0.4) -- (0,0.4) -- (0,0);
\end{scope}
\begin{scope}[xshift=0.8cm,yshift=1.2cm]
\draw[very thin,fill] (0,0) -- (0.4,0) -- (0.4,0.4) -- (0,0.4) -- (0,0);
\end{scope}
\begin{scope}[xshift=0.8cm,yshift=1.6cm]
\draw[very thin,fill] (0,0) -- (0.4,0) -- (0.4,0.4) -- (0,0.4) -- (0,0);
\end{scope}
\begin{scope}[xshift=1.2cm,yshift=1.6cm]
\draw[very thin,fill] (0,0) -- (0.4,0) -- (0.4,0.4) -- (0,0.4) -- (0,0);
\end{scope}
\begin{scope}[xshift=1.6cm,yshift=1.6cm]
\draw[very thin,fill] (0,0) -- (0.4,0) -- (0.4,0.4) -- (0,0.4) -- (0,0);
\end{scope}
\begin{scope}[xshift=2cm,yshift=1.6cm]
\draw[very thin,fill] (0,0) -- (0.3,0) -- (0.3,0.4) -- (0,0.4) -- (0,0);
\end{scope}
\draw[->,very thick] (-2.3,-1.9) to[out=-15,in=145] (2.5,1.7);
\end{scope}
\end{tikzpicture}
\end{center}
\caption{\small The classical limit where the wave packet is much narrower than the size of the cells the sequence of quantum microstates associated to cells in phase space just follows the expectation values of $x$ and $p$ which Ehrenfest's theorem implies is the classical trajectory.}
\label{f3}
\end{figure}
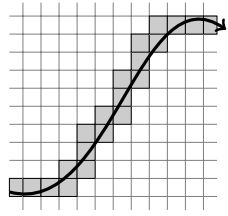

\subsection{Including the momentum}\label{s6.4}

The simple model that we have described can only be partly correct because we have avoided any mention of the particle's momentum. Classically the particles state is specified by both a position and momentum and so we need to consider enlarging $\EA$ to include coarse grained momentum operators.\footnote{In fact, without the momentum it would be possible for two apparently decoherent wave packets that cross in space to interfere. This is because classical trajectories in configuration space can intersect. But once momentum is includes the trajectories in phase space cannot intersect and these kind of interferences are excluded.} 

The wave function of a macroscopic object is expected to be narrow in 
both position and momentum space compared with the resolution scales $\delta_x$ and $\delta_p$ and so 
there is only one quantum microstate, the original state itself $\ket{\Psi}$, with a probability close to one.
In this classical limit, the stochastic process becomes essentially deterministic and the stochastic trajectory localises around a classical trajectory in phase space as illustrated in figure \ref{f3}.

\section{Quantum Measurement}\label{s6.5}

In this section, we consider a very simple measuring scenario to see how the formalism can lead to definite macroscopically distinct outcomes with probabilities that are given by Born's rule (at least in the case of an efficient measuring device).

In the model, the total system includes a microscopic system that has an observable $A$ that we want to measure. Let $\ket{\psi_a}$ be the eigenstates of $A$ and the initial state of the microscopic system is $\sum_a\lambda_a\ket{\xi_a}$.
The system is coupled to a measuring device with a Hamiltonian that ensures an evolution of the combined system of the form\footnote{In the following, we have assumed that the measuring device is 100\% efficient. Making the generalization to a more realistic measuring device does not change the central conclusion that we come to below that definite outcomes are obtained. However, the resulting probabilities will no longer be equal to $|\lambda_a|^2$. So the derivation of the usual Born rule relies on the fact that the measuring device is 100 \% efficient.}
\EQ{
\Big[\sum_a\lambda_a\ket{\xi_a}\Big]\otimes\ket{\psi_0}\longrightarrow
\sum_a\lambda_a\ket{\xi_a(t)}\otimes\ket{\psi_a(t)}\ .
\label{pp3}
}
Note that in general the states of the microscopic system will be changed as a result of the interaction with the measuring device. The states $\ket{\xi_a(t)}$ remain normalized but are not necessarily orthogonal.

The idea is that the states of the measuring device $\ket{\psi_a(t)}$ become macroscopically distinct for $t\geq T$. After this time a measurement could be said to have occurred. In our simple model, we will suppose that the measuring device consists of a pointer with a position. So the states of the measuring device that are entangled with microscopic system are wave functions $\psi_a(x,t)$. The idea is that the measuring device is designed in such a way that for $t>T$, each component $\psi_a(x,t)$ is a narrow wave packet peaked around a distinct position $x^{(a)}$. Note that each component $\psi_a(x,t)$ is separately normalized.

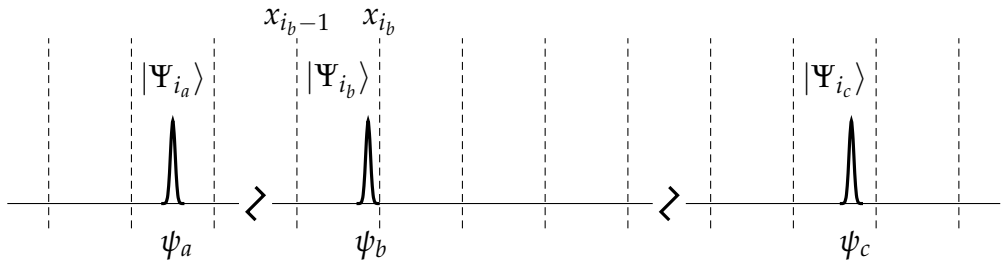
\begin{figure}[ht]
\begin{center}
\begin{tikzpicture}[scale=1.1]
\draw[very thick,-] (1.5,0.2) -- (1.6,0.1) -- (1.4,-0.1) -- (1.5,-0.2);
\draw[very thick,-] (6.5,0.2) -- (6.6,0.1) -- (6.4,-0.1) -- (6.5,-0.2);
\draw[-] (-1.5,0) -- (1.3,0);
\draw[-] (1.7,0) -- (6.3,0);
\draw[-] (6.7,0) -- (10.5,0);
\node at (0.5,1.5) (a1) {$\ket{\Psi_{i_a}}$};
\node at (2.5,1.5) (a1) {$\ket{\Psi_{i_b}}$};
\node at (8.5,1.5) (a1) {$\ket{\Psi_{i_c}}$};
\draw[densely dashed] (-1,-0.3) -- (-1,2);
\draw[densely dashed] (0,-0.3) -- (0,2);
\draw[densely dashed] (1,-0.3) -- (1,2);
\draw[densely dashed] (2,-0.3) -- (2,2);
\draw[densely dashed] (3,-0.3) -- (3,2);
\draw[densely dashed] (4,-0.3) -- (4,2);
\draw[densely dashed] (5,-0.3) -- (5,2);
\draw[densely dashed] (6,-0.3) -- (6,2);
\draw[densely dashed] (7,-0.3) -- (7,2);
\draw[densely dashed] (8,-0.3) -- (8,2);
\draw[densely dashed] (9,-0.3) -- (9,2);
\draw[densely dashed] (10,-0.3) -- (10,2);
\begin{scope}[xscale=0.02,yscale=1,xshift=25cm,yshift=0cm]
\draw[-,very thick] plot[smooth] coordinates {(-7, 0) (-6, 0) (-5., 0.00193045) (-4., 
  0.0183156) (-3., 0.105399) (-2., 0.367879) (-1., 0.778801) (0., 
  1.) (1., 0.778801) (2., 0.367879) (3., 0.105399) (4., 
  0.0183156) (5., 0.00193045) (6., 0) (7., 0)};
\end{scope}
\begin{scope}[xscale=0.02,yscale=1,xshift=143cm,yshift=0cm]
\draw[-,very thick] plot[smooth] coordinates {(-7, 0) (-6, 0) (-5., 0.00193045) (-4., 
  0.0183156) (-3., 0.105399) (-2., 0.367879) (-1., 0.778801) (0., 
  1.) (1., 0.778801) (2., 0.367879) (3., 0.105399) (4., 
  0.0183156) (5., 0.00193045) (6., 0) (7., 0)};
\end{scope}
\begin{scope}[xscale=0.02,yscale=1,xshift=435cm,yshift=0cm]
\draw[-,very thick] plot[smooth] coordinates {(-7, 0) (-6, 0) (-5., 0.00193045) (-4., 
  0.0183156) (-3., 0.105399) (-2., 0.367879) (-1., 0.778801) (0., 
  1.) (1., 0.778801) (2., 0.367879) (3., 0.105399) (4., 
  0.0183156) (5., 0.00193045) (6., 0) (7., 0)};
\end{scope}
\node at (0.55,-0.5) (a1) {$\psi_a$};
\node at (2.9,-0.5) (a1) {$\psi_b$};
\node at (8.75,-0.5) (a1) {$\psi_c$};
\node at (2,2.2) (a1) {$x_{i_b-1}$};
\node at (3,2.2) (a1) {$x_{i_b}$};
\end{tikzpicture}
\end{center}
\caption{\small The measurement scenario. In the final state, each measurement outcome corresponds to a distinct quantum microstate which are ergodically disjoint.}
\label{f2}
\end{figure}

If we now apply our interpretation by using the coarse-grained position operators $\EA=\{\Pi^{(x)}_i\}$ as in section \ref{s6}. At some time $t>T$, the quantum microstates are
\EQ{
\ket{\Psi_i}=\frac1{c_i}\int_{x_{i-1}}^{x_i}dx\,\sum_a\lambda_a\ket{\xi_a(t)}\otimes\psi_a(x,t)\ket{x}\ ,
}
with
\EQ{
|c_i|^2=\int_{x_{i-1}}^{x_i}dx\,\sum_{ab}\lambda_a^*\lambda_b\bra{\xi_a(t)}\xi_b(t)\rangle\psi_a(x,t)^*\psi_b(x,t)\ .
\label{lss}
}

For the measuring device to be accurate, we need its spatial resolution scale $\delta_x$ to be much smaller than the separation of $x^{(a)}$. It is also realistic to assume that the spatial resolution scale $\delta_x$ is much larger than the width of the individual wave packets $\psi_a(x,t)$. Given these assumptions, it follows that $|c_i|^2$ is only appreciable if some $x^{(a)}\in[x_{i-1},x_i]$, for some particular $i$ which we can call $i_a$. So only the diagonal term $a=b$ contributes in \eqref{lss}. Note that only one $x^{(a)}$ can lie in any interval in which case, it follows for this interval that
\EQ{
|c_{i_a}|^2\approx|\lambda_a|^2\ .
\label{rr2}
}
Since the initial quantum microstate is unique, $|\lambda_a|^2$ represents the probability that the quantum microstate is $\ket{\Psi_{i_a}}$ at time $t>T$. This amounts to a derivation of the usual Born rule. The matching of the components of the underlying state to the quantum microstates is illustrated in figure \ref{f2}.

\begin{figure}[ht]
\begin{center}
\begin{tikzpicture}[scale=0.6,fill=black!20]
\draw[step=.4cm,gray,very thin] (-2.3,-2.3) grid (2.3,2.3);
\begin{scope}[xshift=-2.3cm,yshift=-2.0cm]
\draw[very thin,fill] (0,0) -- (0.3,0) -- (0.3,0.4) -- (0,0.4) -- (0,0);
\end{scope}
\begin{scope}[xshift=-2.0cm,yshift=-2.0cm]
\draw[very thin,fill] (0,0) -- (0.4,0) -- (0.4,0.4) -- (0,0.4) -- (0,0);
\end{scope}
\begin{scope}[xshift=-2cm,yshift=-1.6cm]
\draw[very thin,fill] (0,0) -- (0.4,0) -- (0.4,0.4) -- (0,0.4) -- (0,0);
\end{scope}
\begin{scope}[xshift=-2cm,yshift=-1.2cm]
\draw[very thin,fill] (0,0) -- (0.4,0) -- (0.4,0.4) -- (0,0.4) -- (0,0);
\end{scope}
\begin{scope}[xshift=-1.6cm,yshift=-1.2cm]
\draw[very thin,fill] (0,0) -- (0.4,0) -- (0.4,0.4) -- (0,0.4) -- (0,0);
\end{scope}
\begin{scope}[xshift=-1.2cm,yshift=-1.2cm]
\draw[very thin,fill] (0,0) -- (0.4,0) -- (0.4,0.4) -- (0,0.4) -- (0,0);
\end{scope}
\begin{scope}[xshift=-0.8cm,yshift=-1.2cm]
\draw[very thin,fill] (0,0) -- (0.4,0) -- (0.4,0.4) -- (0,0.4) -- (0,0);
\end{scope}
\begin{scope}[xshift=-0.8cm,yshift=-0.8cm]
\draw[very thin,fill] (0,0) -- (0.4,0) -- (0.4,0.4) -- (0,0.4) -- (0,0);
\end{scope}
\begin{scope}[xshift=-0.4cm,yshift=-0.8cm]
\draw[very thin,fill] (0,0) -- (0.4,0) -- (0.4,0.4) -- (0,0.4) -- (0,0);
\end{scope}
\begin{scope}[xshift=-0.8cm,yshift=-0.8cm]
\draw[very thin,fill] (0,0) -- (0.4,0) -- (0.4,0.4) -- (0,0.4) -- (0,0);
\end{scope}
\begin{scope}[xshift=-0cm,yshift=-0.4cm]
\draw[very thin,fill] (0,0) -- (0.4,0) -- (0.4,0.4) -- (0,0.4) -- (0,0);
\end{scope}
\begin{scope}[xshift=-0cm,yshift=-0cm]
\draw[very thin,fill] (0,0) -- (0.4,0) -- (0.4,0.4) -- (0,0.4) -- (0,0);
\end{scope}
\begin{scope}[xshift=-0.4cm,yshift=-0.4cm]
\draw[very thin,fill] (0,0) -- (0.4,0) -- (0.4,0.4) -- (0,0.4) -- (0,0);
\end{scope}
\begin{scope}[xshift=0.4cm,yshift=0cm]
\draw[very thin,fill] (0,0) -- (0.4,0) -- (0.4,0.4) -- (0,0.4) -- (0,0);
\end{scope}
\begin{scope}[xshift=0.4cm,yshift=0.4cm]
\draw[very thin,fill] (0,0) -- (0.4,0) -- (0.4,0.4) -- (0,0.4) -- (0,0);
\end{scope}
\begin{scope}[xshift=0.4cm,yshift=0.8cm]
\draw[very thin,fill] (0,0) -- (0.4,0) -- (0.4,0.4) -- (0,0.4) -- (0,0);
\end{scope}
\begin{scope}[xshift=0.8cm,yshift=0.8cm]
\draw[very thin,fill] (0,0) -- (0.4,0) -- (0.4,0.4) -- (0,0.4) -- (0,0);
\end{scope}
\begin{scope}[xshift=0.8cm,yshift=1.2cm]
\draw[very thin,fill] (0,0) -- (0.4,0) -- (0.4,0.4) -- (0,0.4) -- (0,0);
\end{scope}
\begin{scope}[xshift=1.2cm,yshift=1.2cm]
\draw[very thin,fill] (0,0) -- (0.4,0) -- (0.4,0.4) -- (0,0.4) -- (0,0);
\end{scope}
\begin{scope}[xshift=1.2cm,yshift=1.6cm]
\draw[very thin,fill] (0,0) -- (0.4,0) -- (0.4,0.4) -- (0,0.4) -- (0,0);
\end{scope}
\begin{scope}[xshift=1.6cm,yshift=1.6cm]
\draw[very thin,fill] (0,0) -- (0.4,0) -- (0.4,0.4) -- (0,0.4) -- (0,0);
\end{scope}
\begin{scope}[xshift=2cm,yshift=1.6cm]
\draw[very thin,fill] (0,0) -- (0.3,0) -- (0.3,0.4) -- (0,0.4) -- (0,0);
\end{scope}
\begin{scope}[xshift=-0.8cm,yshift=-1.6cm]
\draw[very thin,fill] (0,0) -- (0.4,0) -- (0.4,0.4) -- (0,0.4) -- (0,0);
\end{scope}
\begin{scope}[xshift=-0.4cm,yshift=-1.6cm]
\draw[very thin,fill] (0,0) -- (0.4,0) -- (0.4,0.4) -- (0,0.4) -- (0,0);
\end{scope}
\begin{scope}[xshift=0cm,yshift=-1.6cm]
\draw[very thin,fill] (0,0) -- (0.4,0) -- (0.4,0.4) -- (0,0.4) -- (0,0);
\end{scope}
\begin{scope}[xshift=0.4cm,yshift=-1.6cm]
\draw[very thin,fill] (0,0) -- (0.4,0) -- (0.4,0.4) -- (0,0.4) -- (0,0);
\end{scope}
\begin{scope}[xshift=0.8cm,yshift=-1.6cm]
\draw[very thin,fill] (0,0) -- (0.4,0) -- (0.4,0.4) -- (0,0.4) -- (0,0);
\end{scope}
\begin{scope}[xshift=1.2cm,yshift=-1.6cm]
\draw[very thin,fill] (0,0) -- (0.4,0) -- (0.4,0.4) -- (0,0.4) -- (0,0);
\end{scope}
\begin{scope}[xshift=1.6cm,yshift=-1.6cm]
\draw[very thin,fill] (0,0) -- (0.4,0) -- (0.4,0.4) -- (0,0.4) -- (0,0);
\end{scope}
\begin{scope}[xshift=2cm,yshift=-1.6cm]
\draw[very thin,fill] (0,0) -- (0.3,0) -- (0.3,0.4) -- (0,0.4) -- (0,0);
\end{scope}
\begin{scope}[xshift=-1.2cm,yshift=-0.8cm]
\draw[very thin,fill] (0,0) -- (0.4,0) -- (0.4,0.4) -- (0,0.4) -- (0,0);
\end{scope}
\begin{scope}[xshift=-1.2cm,yshift=-0.4cm]
\draw[very thin,fill] (0,0) -- (0.4,0) -- (0.4,0.4) -- (0,0.4) -- (0,0);
\end{scope}
\begin{scope}[xshift=-1.2cm,yshift=0cm]
\draw[very thin,fill] (0,0) -- (0.4,0) -- (0.4,0.4) -- (0,0.4) -- (0,0);
\end{scope}
\begin{scope}[xshift=-1.2cm,yshift=0.4cm]
\draw[very thin,fill] (0,0) -- (0.4,0) -- (0.4,0.4) -- (0,0.4) -- (0,0);
\end{scope}
\begin{scope}[xshift=-1.2cm,yshift=0.8cm]
\draw[very thin,fill] (0,0) -- (0.4,0) -- (0.4,0.4) -- (0,0.4) -- (0,0);
\end{scope}
\begin{scope}[xshift=-1.6cm,yshift=0.8cm]
\draw[very thin,fill] (0,0) -- (0.4,0) -- (0.4,0.4) -- (0,0.4) -- (0,0);
\end{scope}
\begin{scope}[xshift=-1.6cm,yshift=1.2cm]
\draw[very thin,fill] (0,0) -- (0.4,0) -- (0.4,0.4) -- (0,0.4) -- (0,0);
\end{scope}
\begin{scope}[xshift=-1.6cm,yshift=1.6cm]
\draw[very thin,fill] (0,0) -- (0.4,0) -- (0.4,0.4) -- (0,0.4) -- (0,0);
\end{scope}
\begin{scope}[xshift=-2cm,yshift=1.6cm]
\draw[very thin,fill] (0,0) -- (0.4,0) -- (0.4,0.4) -- (0,0.4) -- (0,0);
\end{scope}
\begin{scope}[xshift=-2cm,yshift=2cm]
\draw[very thin,fill] (0,0) -- (0.4,0) -- (0.4,0.3) -- (0,0.3) -- (0,0);
\end{scope}
\draw[->,very thick] (-2.3,-1.9) to[out=15,in=165] (-1.1,-0.98);
\draw[->,very thick] (-1.1,-0.98) to[out=70,in=290] (-1.9,2.5);
\draw[->,very thick] (-1.1,-0.98) to[out=10,in=170] (2.5,1.8);
\draw[->,very thick] (-1.1,-0.98) to[out=-30,in=-170] (2.5,-1.3);
\node at (-5.8,0.8) (g1) {interaction with};
 \node at (-5.8,0) (g3) {microscopic system};
\node at (5,0) (g2) {ergodically disjoint}; 
\draw[->] (g3) -- (-1.3,-0.88);
\draw[->] (g2) -- (2,1.5);
\draw[->] (g2) -- (2,-1.1);
\end{tikzpicture}
\end{center}
\caption{\small A picture of a quantum measurement in the phase space of a very simple measuring device with one degree of freedom. The system follows a coarse graining of the classical trajectory. At the moment of interaction with the microscopic system, the trajectory splits into several distinct trajectories that soon become ergodically disjoint. The measurement outcome is determined by which of the disjoint trajectories the system ends up in.}
\label{f4}
\end{figure}
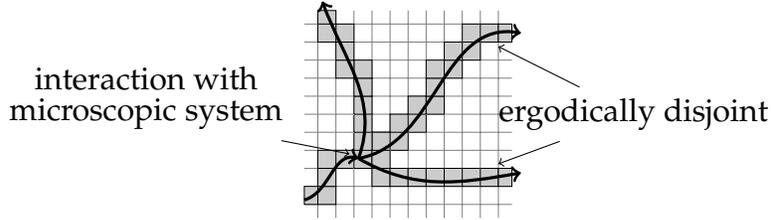

The last important point is that once the system is in the quantum microstate $\ket{\Psi_i}$ there is only a minute probability for making a transition to a macroscopically distinct state $\ket{\Psi_j}$. So it is the breaking of ergodicity that leads to distinct classical outcomes and hence solves the measurement problem. The situation is illustrated in figure \ref{f4}.
At the end of the measurement, from the point of view of $\EA$ it is prudent to reduce the underlying state vector to the component 
$\ket{\xi_a(t)}\otimes\psi_a(x,t)$ that is actually realized as a quantum microstate.

\section{EPR}\label{s9}

In this section, we give an account of Bohm's version 
\cite{Bohm:1951qt,BohmAharonov:1957depperp} of the 
classic thought experiment of Einstein, Podolsky, and Rosen (EPR) \cite{EinsteinPodolskyRosen:1935cqmdprbcc}. Anticipating the result, we find that is no non-locality or breakdown of causality, the experiment simply reveals the correlation of an entangled state of two qubits.
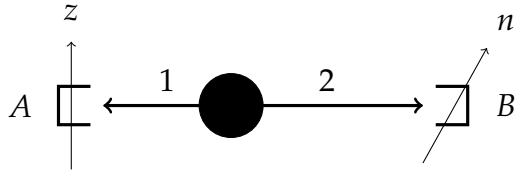
\begin{figure}[ht]
\begin{center}
\begin{tikzpicture}[scale=0.85]
\node at (4.3,0) (a1) {$B$};
\node at (-3.3,0) (a1) {$A$};
\node at (1.5,0.4) (a1) {$2$};
\node at (-1,0.4) (a1) {$1$};
\filldraw[black] (0,0) circle (0.5cm);
\draw[->,very thick] (0,0) -- (3,0);
\draw[->,very thick] (0,0) -- (-2,0);
\begin{scope}[xshift=3cm]
\draw[very thick] (0.2,0.3) -- (0.7,0.3) -- (0.7,-0.3) -- (0.2,-0.3);
\draw[->] (0,-0.9) -- (1,0.9);
\node at (1.3,1.3) (b1) {$n$};
\end{scope}
\begin{scope}[xshift=-2cm,rotate=180]
\draw[very thick] (0.2,0.3) -- (0.7,0.3) -- (0.7,-0.3) -- (0.2,-0.3);
\draw[<-] (0.5,-1) -- (0.5,1);
\node at (0.5,-1.45) (b1) {$z$};
\end{scope}
\end{tikzpicture}
\end{center}
\caption{\small The EPR-Bohm thought experiment. Two qubits in the entangled state $\ket{\Phi}$ are produced at the source and then recoil back-to-back towards 2 qubit detectors $A$ and $B$ designed to measure the component of the spin along directions $z$ and $n$, respectively. In our set up, we choose an inertial frame for which the interaction between $A$ and 1 happens before $B$ and 2.}
\label{f18}
\end{figure}

In order to describe spin measurements on the qubits, we introduce a very simple measuring device with three quantum microstates $\ket{A_0}$ and $\ket{A_\pm}$. In this simple model the quantum microstates are actually the possible macrostates of the measuring device; in other words, in order to focus on the essential details, we avoid the fact that, in reality, the measuring device is a complicated thermodynamic object with many degrees of freedom. The measuring device is designed so that the solution of the Schr\"odinger equation for its interaction with a qubit from an initial time $t=0$ to a final time $T$ is of the form
\EQ{
\ket{A_0}\otimes\big(c_+\ket{z^+}+c_-\ket{z^-}\big)\longrightarrow
c_+\ket{A_+}\otimes\ket{z^+}+c_-\ket{A_-}\otimes\ket{z^-}\ .
}
The stochastic process associated to the observables of the measuring device yield integrated probabilities
\EQ{
p_{+|0}(T,0)=|c_+|^2\ ,\qquad p_{-|0}(T,0)=|c_-|^2\ .
}
In this simple model, we can identify the quantum microstates in the final state with the reduced state vectors of the Copenhagen interpretation. In a realistic measuring device, the reduced state vector only emerges when the ergodic components in \eqref{bb2} become identifiable. This happens in a macroscopically short time but is ultimately a dynamical question.

The experiment is described in figure \ref{f18} with the qubits being produced in the singlet state
\EQ{
\ket{\Phi}=\frac1{\sqrt2}\big(\ket{z^+z^-}-\ket{z^-z^+}\big)
}
and so the initial state of the overall system is
\EQ{
\ket{\Psi(t_1)}=\ket{A_0 B_0}\otimes\ket{\Phi}\ .
\label{sw1}
}
In a certain inertial frame, at time $t_A>t_1$, the measuring device $A$, set to measure the $z$ component of the spin, interacts with qubit 1. After a short time, the state of qubit $1$ becomes entangled with $A$ and the state of the total system becomes, for $t_2>t_A$,
\EQ{
\ket{\Psi(t_2)}=\frac1{\sqrt2}\big(\ket{A_+B_0z^+z^-}-\ket{A_-B_0z^-z^+}\big)\ .
}
Then at a time $t_B$ the measuring device $B$, set to measure the component of the spin along a vector $n$ at an angle $2\theta$ to the $z$ axis, interacts with qubit 2 
producing, for $t_3>t_B$, a state\footnote{The relation between the basis of qubit states along the $z$ and $n$ axes is
$\ket{z^+}=\cos\theta \ket{n^+}-\sin\theta \ket{n^-}$ and $\ket{z^-}=\sin\theta \ket{n^+}+\cos\theta \ket{n^-}$.}
\EQ{
\ket{\Psi(t_3)}&=\frac1{\sqrt2}\Big(\sin\theta \ket{A_+B_+z^+n^+}+\cos\theta \ket{A_+B_-z^+n^-}\\
&\qquad\qquad-\cos\theta \ket{A_-B_+z^-n^+}+\sin\theta \ket{A_-B_-z^-n^-}\Big)\ .
}
Just to emphasize, the states $\ket{\Psi(t_i)}$, $i=1,2,3$, follow from solving the Schr\"odinger equation of the system with a Hamiltonian that only has local interaction terms between the measuring devices and their associated qubit, $A$ with $1$ and $B$ with $2$.

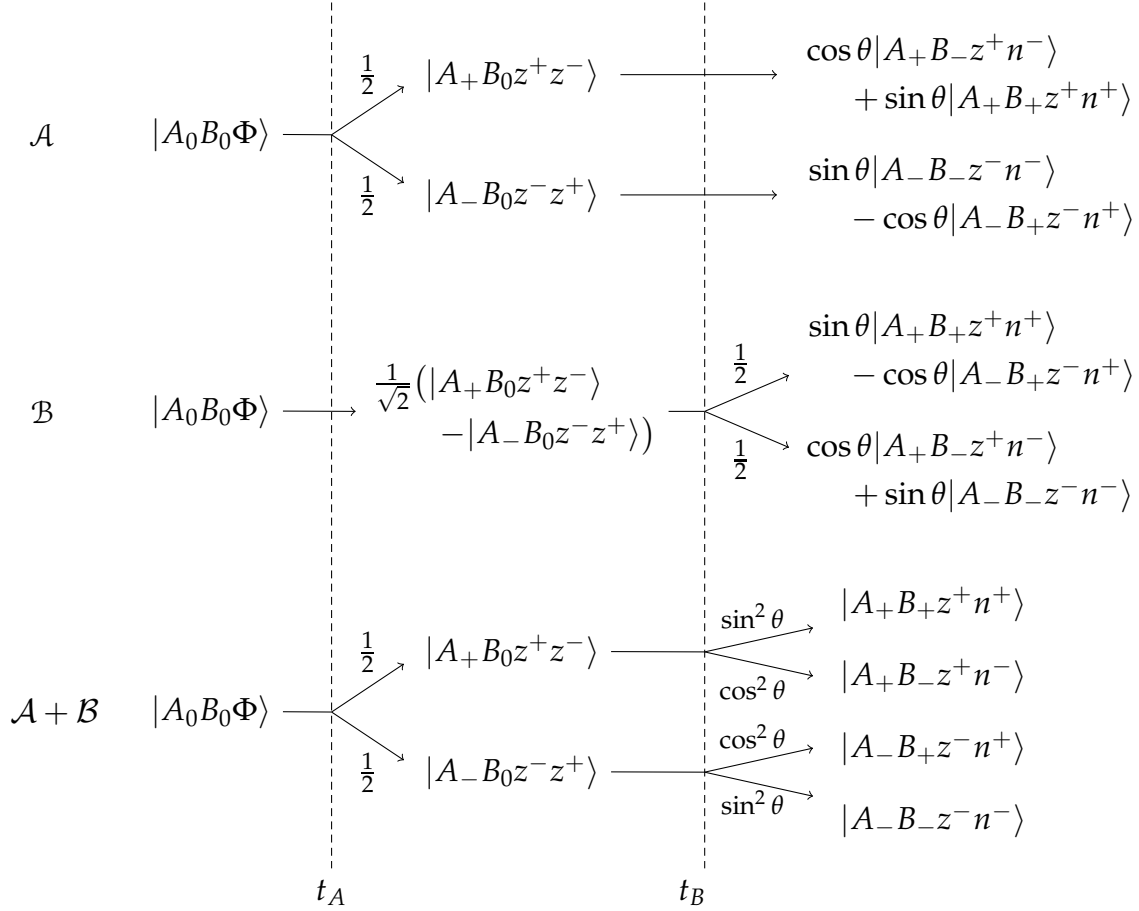
\begin{figure}[ht]
\begin{center}
\begin{tikzpicture}[xscale=1.6,yscale=1.6]
\begin{scope}[yshift=-2cm]
\node at (0.6,10) (b1) {${\EuScript A}$};
\node at (2,10) (b2) {$\ket{A_0B_0\Phi}$};
\node at (4.5,10.5) (b3) {$\ket{A_+B_0z^+z^-}$};
\node at (4.5,9.5) (b4) {$\ket{A_-B_0z^-z^+}$};
\node at (8,10.7) (b5) {$\cos\theta \ket{A_+B_-z^+n^-}$};
\node at (8.5,10.3) (b6) {$+\sin\theta \ket{A_+B_+z^+n^+}$};
\node at (8,9.7) (b7) {$\sin\theta \ket{A_-B_-z^-n^-}$};
\node at (8.5,9.3) (b8) {$-\cos\theta \ket{A_-B_+z^-n^+}$};
\draw[-] (b2) -- (3,10);
\draw[->] (3,10) -- (3.6,10.4);
\draw[->] (3,10) -- (3.6,9.6);
\draw[->] (5.4,10.5) -- (6.7,10.5);
\draw[->] (5.4,9.5) -- (6.7,9.5);
\node at (3.3,10.5) (z1) {$\frac12$};
\node at (3.3,9.5) (z1) {$\frac12$};
\end{scope}
\begin{scope}[yshift=-4.3cm]
\node at (0.6,10) (b1) {${\EuScript B}$};
\node at (2,10) (b2) {$\ket{A_0B_0\Phi}$};
\node at (4.3,10.2) (b3) {$\frac1{\sqrt2}\big(\ket{A_+B_0z^+z^-}$};
\node at (4.8,9.8) (b4) {$-\ket{A_-B_0z^-z^+}\big)$};
\node at (8,10.7) (b5) {$\sin\theta \ket{A_+B_+z^+n^+}$};
\node at (8.5,10.3) (b6) {$-\cos\theta \ket{A_-B_+z^-n^+}$};
\node at (8,9.7) (b7) {$\cos\theta \ket{A_+B_-z^+n^-}$};
\node at (8.5,9.3) (b8) {$+\sin\theta \ket{A_-B_-z^-n^-}$};
\draw[->] (b2) -- (3.2,10);
\draw[-] (5.8,10) -- (6.1,10);
\draw[->] (6.1,10) -- (6.8,10.3);
\draw[->] (6.1,10) -- (6.8,9.7);
\node at (6.4,10.4) (z1) {$\frac12$};
\node at (6.4,9.6) (z1) {$\frac12$};
\end{scope}
\begin{scope}[yshift=-6.8cm]
\node at (0.7,10) (b1) {$\cal{A+B}$};
\node at (2,10) (b2) {$\ket{A_0B_0\Phi}$};
\node at (4.5,10.5) (b3) {$\ket{A_+B_0z^+z^-}$};
\node at (4.5,9.5) (b4) {$\ket{A_-B_0z^-z^+}$};
\node at (8,10.9) (b5) {$\ket{A_+B_+z^+n^+}$};
\node at (8,10.3) (b6) {$\ket{A_+B_-z^+n^-}$};
\node at (8,9.7) (b7) {$\ket{A_-B_+z^-n^+}$};
\node at (8,9.1) (b8) {$\ket{A_-B_-z^-n^-}$};
\draw[-] (b2) -- (3,10);
\draw[->] (3,10) -- (3.6,10.4);
\draw[->] (3,10) -- (3.6,9.6);
\draw[-] (b3) -- (6.1,10.5);
\draw[->] (6.1,10.5) -- (7,10.7);
\draw[->] (6.1,10.5) -- (7,10.3);
\draw[-] (b4) -- (6.1,9.5);
\draw[->] (6.1,9.5) -- (7,9.7);
\draw[->] (6.1,9.5) -- (7,9.3);
\node at (3.3,10.5) (z1) {$\frac12$};
\node at (3.3,9.5) (z1) {$\frac12$};
\node at (6.5,10.8) (z1) {\footnotesize$\sin^2\theta$};
\node at (6.5,10.2) (z1) {\footnotesize$\cos^2\theta$};
\node at (6.5,9.8) (z1) {\footnotesize$\cos^2\theta$};
\node at (6.5,9.25) (z1) {\footnotesize$\sin^2\theta$};
\end{scope}
\draw[-,densely dashed] (3,9.1) -- (3,1.9);
\draw[-,densely dashed] (6.1,9.1) -- (6.1,1.9);
\node at (3,1.7) (z1) {$t_A$};
\node at (6,1.7) (z1) {$t_B$};
\end{tikzpicture}
\end{center}
\caption{\small The Markov chains of the quantum microstates (the macrostates or reduced state vectors in this simple model) of the sets of observables ${\EuScript A}$ and ${\EuScript B}$ associated to the two measuring devices and the global view $\cal{A}+{\EuScript B}$ at times $t_1$, $t_2$ and $t_3$ between the measurements at $t_A$ and $t_B$. Note that the quantum microstates of ${\EuScript A}$ and ${\EuScript B}$ always have the factored form $\ket{A_i}\otimes\ket{\psi}$and $\ket{B_i}\otimes\ket{\psi}$, respectively, for $i=0,\pm$ and for some $\ket{\psi}$ in the appropriate complementary subsystem. Also shown are the probabilities of the Markov chains.}
\label{f7}
\end{figure}

In this system, there are three different perspectives associated to the observables of the two measuring devices ${\EuScript A}$ and ${\EuScript B}$ as well as the global view associated to ${\EuScript A}\cup{\EuScript B}$. It is important to appreciate that the descriptions ${\EuScript A}$ and ${\EuScript B}$ cannot answer questions about correlations between the measuring devices, it is only the more refined global view ${\EuScript A}\cup{\EuScript B}$ that can do that. Figure \ref{f7} shows the three Markov processes.

The quantum microstates of ${\EuScript A}$ and ${\EuScript B}$ are related to those of ${\EuScript A}\cup{\EuScript B}$ precisely as in \eqref{pn1}. It is important that the changes in the quantum microstates are driven by local interactions: $A$ with qubit 1 at $t_A$ and $B$ with qubit 2 at $t_B$. It is worth remarking on the fact that the quantum microstates associated to ${\EuScript A}$ and ${\EuScript B}$ are not directly comparable: by focussing on, say, ${\EuScript A}$, we forego any knowledge of ${\EuScript B}$, and vice-versa. The only way to ask about joint properties of the two measuring devices is via the quantum microstates associated to ${\EuScript A}\cup{\EuScript B}$. The quantum  microstates of the latter are always refinements of those of ${\EuScript A}$ and ${\EuScript B}$ as is clear from the last column in figure \ref{f7}. With our simple description of the measuring devices, we can identify the quantum microstates of ${\EuScript A}$ and ${\EuScript B}$ with the reduced state vectors of the Copenhagen interpretation. It then becomes apparent that different local observers, in this case ${\EuScript A}$ and ${\EuScript B}$, perform their own reduction of the state vector to take account the results of their local measurements.\footnote{The fact that each local observer has their own reduced state vector after making local measurements has been emphasised in \cite{Englert}.}

One way to think of the relation between the refined view ${\EuScript A}\cup{\EuScript B}$ and, say, $\EA$ is the following. When the measurement is made at $B$, the quantum microstates
of ${\EuScript A}\cup{\EuScript B}$, for example $\ket{A_+B_0z^+ z^-}$ splits into the pair
$\ket{A_+ B_\pm z^+n^\pm}$. But the set of observables $\EA$ cannot distinguish this pair of states:
\EQ{
\bra{A_+B_i\,z^+n^i}{\cal O}_n\otimes 1_B\otimes 1_1\otimes1_2\ket{A_+B_j\,z^+n^j}=\mu_n\delta_{ij}\ ,\qquad\forall{\cal O}_n\in\EA\ ,
}
it is only observables in ${\EuScript B}$ that can.
Hence, the pair of states do not satisfy the non-degeneracy condition \eqref{sa2} and that is why quantum microstate with respect to $\EA$ is the linear combination
\EQ{
\cos\theta\ket{A_+B_-z^+n^-}+\sin\theta\ket{A_+B_+z^+n^+}\ .
}

The stochastic process associated to the set of observables ${\EuScript A}$ for measuring device $A$ gives the non-vanishing integrated probabilities
\EQ{
p_{+|0}^{{\EuScript A}}(t_2,t_1)=p_{-|0}^{{\EuScript A}}(t_2,t_1)=\frac12\ ,\qquad p_{+|+}^{{\EuScript A}}(t_3,t_2)=p_{-|-}^{{\EuScript A}}(t_3,t_2)=1\ .
}
In particular nothing happens when the measurement is made at $B$. While that associated to ${\EuScript B}$ gives
\EQ{
p_{0|0}^{{\EuScript B}}(t_2,t_1)=1\ ,\qquad p_{+|0}^{{\EuScript B}}(t_3,t_2)=p_{-|0}^{{\EuScript B}}(t_3,t_2)=\frac12\ .
}
In this case, nothing happens when the measurement is made at $A$. Just to be completely clear, in figure \ref{f7}, the change in the quantum microstate of ${\EuScript B}$ from $t_1$ to $t_2$, whilst the measurement is being made at $A$, is just evolution by the Schr\"odinger equation. Likewise is the change in either of the quantum microstates of ${\EuScript A}$ from $t_2$ to $t_3$.

The global view ${\EuScript A}\cup{\EuScript B}$ is provides a more refined description of the system that yields the correlations between measurements at $A$ and $B$:
\EQ{
p_{+0|00}^{{\EuScript A}\cup{\EuScript B}}(t_2,t_1)&=p_{-0|00}^{{\EuScript A}\cup{\EuScript B}}(t_2,t_1)=\frac12\ ,\\
p_{++|+0}^{{\EuScript A}\cup{\EuScript B}}(t_3,t_2)&=p_{--|-0}^{{\EuScript A}\cup{\EuScript B}}(t_3,t_2)=\sin^2\theta\ ,\\
p_{+-|+0}^{{\EuScript A}\cup{\EuScript B}}(t_3,t_2)&=p_{-+|-0}^{{\EuScript A}\cup{\EuScript B}}(t_3,t_2)=\cos^2\theta\ ,
}
giving the overall probabilities
\EQ{
p_{++|00}^{{\EuScript A}\cup{\EuScript B}}(t_3,t_1)&=p_{--|00}^{{\EuScript A}\cup{\EuScript B}}(t_3,t_1)=\frac12\sin^2\theta\ ,\\
p_{-+|00}^{{\EuScript A}\cup{\EuScript B}}(t_3,t_1)&=p_{+-|00}^{{\EuScript A}\cup{\EuScript B}}(t_3,t_1)=\frac12\cos^2\theta\ ,
}
which agree with the probabilities calculated in the Copenhagen interpretation using the Born rule.

So in this account of the EPR-Bohm thought experiment, the stochastic process yields the conventional
quantum mechanical predictions in a way that involves only local interactions between measuring devices and qubits. There is no mystery or spooky action-at-a-distance, on the contrary there is simply the unveiling of an underlying correlation carried by the qubits---albeit of a kind that cannot be mimicked by a classical system.

\section{The Emergence of Classical Statistical Mechanics}\label{s7}

It should be apparent that our interpretation has some important implications for the way that classical SM arises from quantum mechanics. 
The first point, as we discussed in the introduction, is that there are many issues in the theory of classical SM that are not completely settled. Take the issue of probability. Essentially, as reviewed, for example, by Wallace \cite{wallace}, there are two distinct points of view called the {\it inferential\/} and {\it dynamical\/}, sometimes portrayed as the Gibbs view versus the Boltzmann. The former views probabilities as arising from our incomplete knowledge of the microstate given the macroscopic information that we have about a system.
On the other hand, in the dynamical point of view, there is no fundamental need for probabilities because classical mechanics is deterministic. However, one can talk about the relative number of particles moving with such and such velocity and 
the dynamics is complicated and it is not possible to follow individual trajectories of microstates and so one has to talk about averages over suitably long times. In ergodic systems it is possible to link these points of view directly, but ergodicity is problematic to formulate and difficult to prove in general. 

Probabilities, of course, naturally arise in quantum mechanics in the conventional Copenhagen interpretation and Wallace \cite{wallace} makes the point that there is then no
need to introduce probabilities in quantum SM twice. It would be much more natural if the probabilities in SM and the Copenhagen interpretation were essentially one and the same. This is what our formalism achieves with the potential to illuminate and solve the conceptual problems of classical SM. 

In fact, a completely quantum approach to SM has its roots in the work of Schr\"odinger \cite{Sch} and von~Neumann \cite{vN} in the 1920s.\footnote{There is now a large literature but a limited set of references are \cite{PopescuShortWinter:2005fsmeisa,PopescuShortWinter:2006efsm,GoldsteinLebowitzTumulkaZanghi:2006ct,LPSW,Sh,GMM,Reimann1,Reimann2,Gogolin,Srednicki:1995pt,Srednicki2,GLMTZ,RS}.}  We will be particularly interested in the approach of von~Neumann leading to his {\it quantum ergodic theorem\/}. 
The idea is to focus on expectation values of suitable coarse grained observables with the aim of showing that the resulting values behave in the way expected in SM.
One can either consider a system in isolation or one can focus on a small subsystem of a much larger system, corresponding to the microcanonical and canonical situations, respectively.
 
There are two important issues: does a system {\it equilibrate\/} and then if so does it {\it thermalize\/}? The issue of equilibration, concerns whether the expectation values of various coarse grained observables become constant, or at least constant for the vast majority of time after some initial transient regime. Note that this is a non-trivial question because the underlying quantum state of the system certainly does not approach a constant. 
The issue of thermalization is whether the equilibrium state gives expectation values that one would expect of the microcanonical or canonical ensembles. So the properties of equilibration and thermalization are properties of the state of the system with respect to a set of coarse grained observables $\EA=\{{\cal O}_n\}$ rather than the underlying quantum state itself.

Let us consider an isolated system and suppose that it is described by a density operator $\rho(t)$, which includes the possibility of a pure state, and consider an observable ${\cal O}\in\EA$. The idea is that it is not the state $\rho(t)$ but the expectation value $\Tr[{\cal O}\rho(t)]$ that equilibrates in the sense that the long time average of it is very close to $\Tr[{\cal O}\rho_0]$ where $\rho_0$ is the time independent component of $\rho(t)$. The latter is just the diagonal component of $\rho(t)$ in the energy eigenstate basis:\footnote{For simplicity, we assume that there are no degeneracies in the $E_n$ and also in the gaps $E_n-E_m$ which is reasonable in a realistic interacting non-integrable quantum system.}
\EQ{
\rho_0=\sum_np_{n}\ket{n}\bra{n}\ ,\qquad p_n=\bra{n}\rho(t)\ket{n}\ ,
}
where $\ket{n}$ are the energy eigenstates and the $p_n$ are time independent probabilities $\sum_np_n=1$. Since many energy eigenstates are expected to be occupied each with a small probability, roughly $p_n\lesssim 10^{-N}$, where $N$ is the number of degrees of freedom of the system, e.g.~$N={\cal O}(10^{23})$.

There are various bounds that can be established. For example,
one can prove \cite{Reimann1,Sh} that the long time average of the variance of the expectation value $\Tr[{\cal O}\rho(t)]$ about the equilibrium $\Tr[{\cal O}\rho_0]$ is bounded above by a quantity of order\footnote{We refer to the original work for the precise statements.}
\EQ{
\Delta({\cal O})^2\sum_np_n^2\ ,
\label{pp1}
}
where $\Delta({\cal O})$ is the range of eigenvalues of ${\cal O}$.
Since the probabilities $p_n$ are very small, \eqref{pp1} is certainly much smaller than the square of the resolution scale $\delta_{\cal O}^2$. Hence the time average of the expectation value of ${\cal O}$ is captured to high degree of the accuracy by the equilibrium state $\rho_0$.

Another interesting quantity is the faction of time that the system spends away from the equilibrium state as measured by the inequality \cite{Reimann2}
\EQ{
\big|\Tr[{\cal O}\rho(t)]-\Tr[{\cal O}\rho_0]\,\big|>\delta_{\cal O}\ ,
}
where $\delta_{\cal O}$ is the resolution scale defined in section \ref{s3.1}.
It can be shown that this fraction of time is bounded above by a quantity of order
\EQ{
\Big(\frac{\Delta({\cal O})}{\delta_{\cal O}}\Big)^2\underset{n}{\text{max}}\ p_n\ .
}
Although the ratio $\Delta({\cal O})/\delta_{\cal O}$ can be large, it is expected to be dominated by the smallness of the eigenvalues $p_n$ and so the implication is that the system spends the vast majority of time close to the equilibrium state with only occasional excursions away. It is important to emphasize, though, that the equilibrium state is not strictly time independent due to these fluctuations.

The issue of thermalization rests on the extent to what expectation values calculated from the  equilibrium state $\rho_0$ are close to the microcanonical ensemble
\EQ{
\rho_\text{mc}=\frac1N\sum_{E_n\in[E,E+\Delta E]}\ket{n}\bra{n}\ .
}
This is a more refined question that requires additional hypotheses. For instance, it follows if the initial state is pure and has a sharply defined energy in the band $[E,E+\Delta E]$ from von Neumann's quantum ergodic therorem \cite{vN}. More generally, it follows from the {\it eigenstate thermalization hypothesis\/} discussed, for example, in \cite{Srednicki:1995pt,Srednicki2,Reimann1,Reimann2,RS}. There are similar statements that can be made for a sub-system $A$ interacting with a larger system $E$. In this case, it is the expectation values of operators on the  subsystem that equilibrate and can give values that are close to the canonical ensemble \cite{GMM,GoldsteinLebowitzTumulkaZanghi:2006ct,PopescuShortWinter:2006efsm,LPSW,Sh,PopescuShortWinter:2005fsmeisa}.

What is striking about this quantum approach to SM is that it completely cuts classical SM and any mention of microstates or probability out of the picture. So whilst the approach is very successful, there does not seem to be a way to see how classical SM can arise in the classical limit. 

However, there are some parallels between our approach to the Copenhagen interpretation and the quantum approach to SM. Both place the expectation values of a set of realistic coarse grained observables $\EA$ centre stage in the sense that the macroscopic state of the system is determined by both the underlying quantum state {\it and\/} the observables. In particular, in the SM context the underlying quantum state by itself is not the relevant object to measure equilibration and thermalization---it certainly does not have these properties---rather it is the expectation values of the observables $\Tr[{\cal O}_n\rho(t)]$, ${\cal O}_n\in\EA$. 

In our proposal, the expectation values of the observables $\Tr[{\cal O}_n\rho(t)]$ as a function of time should be replaced by the expectation values in the sequence of quantum microstates determined by the stochastic process $\bra{\Psi_{i(t)}}{\cal O}_n\ket{\Psi_{i(t)}}$. The key question for us is:

\vspace{0.2cm}
\begin{changemargin}{0.3cm}{0.3cm}
\vspace{0.2cm}{\it
\no To what extent is the  long time average of  $\bra{\Psi_{i(t)}}{\cal O}_n\ket{\Psi_{i(t)}}$ captured by the microcanonical equilibrium expectation value $\Tr[{\cal O}_n\rho_\text{mc}]$?
\vspace{0.2cm}
}\end{changemargin}

\vspace{0.2cm}\no In order to answer this, let us make us some assumptions:

\begin{enumerate}
\item We will take the overall state to be the pure state $\rho=\ket{\Psi}\bra{\Psi}$ with sharply defined energy
in a microcanonical shell $[E,E+\Delta E]$ defined by a subspace of the Hilbert space $\BH_\text{mc}\subset\BH$ with a large dimension $N$ with $\Delta E$ is macroscopically small but microscopically large so that $N$ is large. 
\item We will assume that the observables ${\cal O}_n$ are explicitly time independent  and commuting, so that the quantum microstates are the simultaneous eigenstates. In addition, none of the ${\cal O}_n$ commutes with the Hamiltonian (except the coarse grained energy: see below) . If the contrary were true, there would be super-selection sectors which would break ergodicity and in that case one would have to restrict to a particular eigenspace. 
\item Finally we assume that the microcanonical 
energy shell defining a subspace of the Hilbert space $\BH_\text{mc}$ has a decomposition in terms of the eigenspaces of the observables 
\EQ{
\BH_\text{mc}=\bigoplus_i\BH_i\ ,\qquad d_i=\text{dim}\,\BH_i\ ,\qquad\sum_id_i=N\ .
}
This seems reasonable because it is natural that $\EA$ contains a coarse grained version of the energy consisting of the projector onto $\BH_\text{mc}$.\footnote{Von~Neumann \cite{vN} calls the family of spaces $\{\BH_i\}$ the ``macro-observer". In our approach we associate the ``macro-observer" to the macroscopic time average of the expectation values $\bra{\Psi_{i(t)}}{\cal O}_n\ket{\Psi_{i(t)}}$.}
\end{enumerate}

When the system, with respect to the set $\EA$, equilibrates in the sense described above, the occupation probabilities of the quantum microstates $|c_i(t)|^2$ also equilibrate. In fact, von~Neumann's quantum ergodic theorem \cite{vN} (see also \cite{GLMTZ}) states that, for any $\ket{\Psi(t)}\in\BH_\text{mc}$ and most times in the long run,
\EQ{
|c_i(t)|^2=\big|\bra{\Psi_i}\Psi(t)\rangle\big|^2=\bra{\Psi(t)}\Pi_i\ket{\Psi(t)}\approx\frac{d_i}{N}\ ,
\label{pps}
}
where $\Pi_i$ is the projector onto $\BH_i$. The exact conditions for the equality are to be found in \cite{vN,GLMTZ} but they include the requirement that the dimensions $d_i$ are suitably large. Note that \eqref{pps} implies
that the expectation values take their microcanonical form:
\EQ{
\bra{\Psi}{\cal O}_n\ket{\Psi}=\sum_i|c_i|^2\bra{\Psi_i}{\cal O}_n\ket{\Psi_i}\approx\Tr[{\cal O}_n\rho_\text{mc}]\ ,
\label{qet}
}
where the first equality is the microscopic Born rule \eqref{sa1}.
The result also follows from the eigenstate thermalization hypothesis \cite{RS}. To emphasize, being in equilibrium does not mean that the $|c_i(t)|^2$ are strictly constant: the equilibrium state allows for fluctuations and very occasional larger excursions away from the average.

So in order to answer the question posed above, we need to establish that the stochastic process in equilibrium is also ergodic so that the long time average of $\bra{\Psi_{i(t)}}{\cal O}_n\ket{\Psi_{i(t)}}$, as in \eqref{tt6}, is equal to the equilibrium values $\Tr[{\cal O}_n\rho_\text{mc}]$. We discussed the issue of ergodicity in section \ref{s4.1}. In equilibrium, the stochastic process is effectively a homogeneous Markov chain for which ergodicity requires that over a long enough time the integrated transition probabilities $p_{i|j}(t+T,t)$ becomes independent of $j$.

The transition rates of the stochastic system are given in \eqref{trt10}.
In order to address the question of ergodicity, we need to investigate 
the matrix elements $\bra{\Psi_j}H\ket{\Psi_i}$, with $i\neq j$.  In an interacting and non-integrable, i.e.~chaotic, quantum system, it is reasonable to expect that the energy basis and the basis of quantum microstates are related by a random unitary matrix. With this hypothesis, one finds approximately that for each pair $i,j$, the non-vanishing transition rate
$T_{ij}$, say, is randomly distributed over a range $[0,d_i\Delta E/(\hbar\sqrt{2N})]$ while $T_{ji}=0$. 

A homogeneous Markov chain with these random transition rates is demonstrably ergodic in numerical simulations. A suitable measure for the time scale for the system to be ergodic is how quickly the system becomes independent its initial state; one finds for large $T$
\EQ{
\big|p_{i|j}(t+T,t)-p_{i|k}(t+T,t)\big|\thicksim\exp\big[-\mu T\big]\ ,
}
where $\mu$ is order $\sqrt N\Delta E/\hbar$. This
shows that the process is ergodic over very short time scales of order $\hbar/(\sqrt N\Delta E)$.
The implication is that the long time average as in \eqref{tt6} over macroscopic time scales $T$, of the expectation values that follow from the stochastic process reproduce those of the microcanonical ensemble:
\EQ{\begin{tikzpicture}
\node at (5,1.1) (c1) {$\dfrac1T\displaystyle\int_0^Tdt\,\bra{\Psi_{i(t)}}{\cal O}_n\ket{\Psi_{i(t)}}\approx\sum_i|c_i|^2\bra{\Psi_i}{\cal O}_n\ket{\Psi_i}=\bra{\Psi}{\cal O}_n\ket{\Psi}\approx\Tr[{\cal O}_n\rho_\text{mc}]\ .$};
\node at (-3,0) (b1) {\phantom{.}};
\node at (2,0) (a1) {ergodicity \eqref{tt6}};
\node at (5.4,-0.8) (a2) {microscopic Born rule \eqref{sa1}};
\node at (9.4,0) (a3) {quantum ergodic theorem \eqref{qet}};
\draw[->] (a1) -- (2.95,1);
\draw[->] (a2) -- (6.85,1);
\draw[->] (a3) -- (9.35,1);
\end{tikzpicture}}

The stochastic dynamics of the expectation values $\bra{\Psi_{i(t)}}{\cal O}_n\ket{\Psi_{i(t)}}$ 
re-instates a dynamical, i.e.~Boltzmann, picture of a thermal ensemble but now in the context of quantum SM. Now we can interpret the density operator as being an ensemble of quantum microstates as in \eqref{sa1} with a clear understanding of the  r\^ole of probability as a fundamental property of the Markov process. So the conceptual picture is clear and the quantum microstates provide a quantum precursor of  the microstates of classical SM.
There remains the program of investigating the stochastic dynamics in detail for particular examples, like a gas of microscopic particles, to find to what extent the stochastic dynamical picture is related to the classical mechanical picture. We have already shown how the latter can arise from the former for a macroscopic particle in section \ref{s6} but one would need to apply the same ideas to the whole gas.

\section{Discussion}\label{s8}

We have described an approach to quantum mechanics which offers the possibility to understand the different deterministic and stochastic elements of the classical limit of quantum mechanics in a more unified way. The key idea is that the underlying quantum system evolves according to the standard rules of quantum mechanics, i.e.~the Schr\"odinger equation, while the stochastic behaviour enters at the level of the set of coarse grained observables that are needed to construct an effective theory that describes the interactions of the system with other macroscopic systems. The central element of this effective theory is a stochastic process that preserves the microscopic Born rule. All the different facets of the classical limit follow from properties of the stochastic process and in particular to what extent the process is ergodic. When ergodicity is broken, the behaviour can be effectively deterministic and the classical limit gives rise to classical mechanics. When the process is ergodic, the behaviour is characteristic of statistical mechanics and we have shown that the quantum system gives a dynamical picture of a thermal ensemble. Finally, when ergodicity is partly broken with large ergodic subsets of states, the formalism provides a natural solution to the measurement problem.

To finish, we can give an account of the fate of Schr\"odinger's famous cat. The evolution of the quantum state leads to a final state vector that involves a linear combination of an alive and a dead cat:
\EQ{
\ket{\Psi(0)}=\ket{\LCATH}\longrightarrow
\ket{\Psi(T)}=c_+\ket{\LCATH}+c_-\ket{\DCATH}\ .
}
In this simple analysis, we need not explicitly describe the other parts of the system, the measuring device, microscopic quantum system and environment: they are assumed to be included in the states $\ket{\LCATH}$ and $\ket{\DCATH}$.

The effective macroscopic description of the cat involves a set of sub-macroscopically coarse grained observables $\EA$. The set of quantum microstates $\{\ket{\Psi_i}\}$ associated to $\EA$ splits into two subsets 
$\{\ket{\Psi_{i_+}}\}$ and $\{\ket{\Psi_{i_-}}\}$,
\EQ{
\ket{\LCATH}=\sum_{i_+}c_{i_+}\ket{\Psi_{i_+}}\ ,\qquad
\ket{\DCATH}=\sum_{i_-}c_{i_-}\ket{\Psi_{i_-}}\ ,
}
corresponding to an alive and a dead cat, respectively. The macroscopic state of the system is then described as the macroscopic time average of the 
expectation values $\bra{\Psi_{i(t)}}{\cal O}_n\ket{\Psi_{i(t)}}$. 
The important point is that for realistic Hamiltonians, the integrated probability over a macroscopic time scale for 
making a transition for a dead to alive, or vice versa, over a macroscopic time scale $t_2-t_1$
\EQ{
p_{i_\pm|j_\mp}(t_2,t_1)\ll1\ .
}
So there is no practical prospect of an alive state $\ket{\Psi_{j_+}}$ making a transition to dead state $\ket{\Psi_{i_-}}$, and vice-versa. So the states of an alive cat and dead cat are ergodically disjoint. Once the quantum microstate of the cat lies in a given ergodic component, which is does with probability $|c_+|^2$ or $|c_-|^2$, it remains there, either dead or alive, with definite
positions for its moving part but also with its more microscopic degrees of freedom in approximate thermal equilibrium.
It makes sense to then remove the ergodically disjoint parts of the state vector that can never be realized, a harmless procedure that is nothing other than the reduction of state vector:
\EQ{
\ket{\Psi(T)}\longrightarrow\ket{\LCATH}\quad\text{or}\quad\ket{\DCATH}\ .
}

\vspace{0.5cm}

\appendix
\appendixpage

\section{Mixed State Generalization}\label{a1}

In this appendix, we show how to generalise the formalism to the case when the state is mixed described by a density matrix $\rho$. We will also focus on the realistic case with a commuting set of observables $\EA$.

As in the case of a pure underlying state, the quantum microstates are just the simultaneous eigenstates of the set $\EA$. The underlying mixed quantum state can always be expressed as 
\EQ{
\rho=\sum_{ij}A_{ij}\rho_{ij}\ ,\qquad \sum_iA_{ii}=1\ ,
\label{ue4}
}
where $\rho_{ij}$ is an operator which maps $\BH_j\to\BH_i$ with $\Tr\rho_{ij}=\delta_{ij}$. 
The microscopic Born rule in this case is expressed as
\EQ{
\Tr\big({\cal O}_n\rho\big)=\sum_iA_{ii}\Tr\big(\rho_{ii}{\cal O}_n\big)\ ,\qquad\forall\ {\cal O}_n\in\EA\ .
}
The r\^ole of the quantum microstate is now played by the diagonal elements $\rho_{ii}$ and  $A_{ii}$ is the generalization of the probability $|c_i|^2$. 

We now define the stochastic process by the same logic as in \eqref{ue1}. Using von Neumann's equation of motion for the density operator, one finds
\EQ{
\frac{dA_{ii}}{dt}=\frac2\hbar\sum_{j\neq i}\IM\Big[A_{ij}\Tr_i\Big([H,\rho_{ij}]-i\hbar\frac{d\rho_{ij}}{dt}\Big)\Big]\ ,
}
where the trace is over $\BH_i$. This yields the transition rates
\EQ{
T_{ij}=\text{max}\Big(\frac2\hbar\IM\Big[\frac{A_{ij}}{A_{jj}}\Tr_i\Big\{[H,\rho_{ij}]-i\hbar\frac{d\rho_{ij}}{dt}\Big\}\Big],0\Big)\ .&
\label{trt13}
}
In the case of a pure state $A_{ij}=c_ic_j^*$ and $\rho_{ij}=\ket{\Psi_i}\bra{\Psi_j}$ we recover the expression \eqref{eqj} using, for $i\neq j$,
\EQ{
\Tr_i\Big\{[H,\rho_{ij}]-i\hbar\frac{d\rho_{ij}}{dt}\Big\}=-\bra{\Psi_j}\Big(H-i\hbar\frac d{dt}\Big)\ket{\Psi_i}\ .
}

If our quantum system is in a mixed state $\rho$ and is interacting with a larger quantum system then the time evolution of $\rho$ will be non-trivial. It would be interesting to generalize our formalism to this case, but we will leave such an endeavour to future work.

\section{Non-Abelian Sets of Observables}\label{a3}

In this appendix, we consider several examples where the set $\EA$ contains operators that do not commute.

\vspace{0.2cm}
\noindent{\bf Example 1:} 
Suppose that $\EA$ consists of the complete set of observables on the finite dimensional Hilbert space $\BH$. If the dimension of $\BH$ is $n$ then the complete set of observables has dimension $n^2$. It quickly becomes apparent in this case that the set of quantum microstates states $\MS$ consists of just $\ket{\Psi}$ itself. In an intuitive sense, if we know the expectation values of all the observables on $\BH$ then we have complete knowledge of the quantum state and so the coarse grained density operator must be equal to $\rho=\ket{\Psi}\bra{\Psi}$. The stochastic process will be trivial and the evolution of the system with respect to $\EA$ is just the same as the deterministic microscopic dynamics encoded in the Schr\"odinger equation. 

So an omnipotent observer $\EA$ sees the system evolve according to the Schr\"odinger equation.

\vspace{0.2cm}
\noindent{\bf Example 2:} 
Suppose that $\EA$ consists of the complete set of observables which as operators act on a factor $\BH_A$ of a tensor product of finite dimensional Hilbert spaces $\BH=\BH_A\otimes\BH_E$. These operators act as ${\cal O}_n\otimes 1_E$. The dimension of $\EA$ is equal to $(\text{dim}\,\BH_A)^2$. 

In this example, the quantum microstates are just the conventional Schmidt states
\EQ{
\ket{\Psi}=\sum_{i=1}^Nc_i\ket{\Psi_i}\ ,\qquad\ket{\Psi_i}=\ket{\psi_i}\otimes\ket{\tilde\psi_i}\ ,
\label{pf1}
} 
where $\bra{\psi_i}\psi_j\rangle=\bra{\tilde\psi_i}\tilde\psi_j\rangle=\delta_{ij}$. The states $\ket{\psi_i}$ and $\ket{\tilde\psi_i}$ are eigenstates of the reduced density matrices 
$\rho^{(A)}=\Tr_E\ket{\Psi}\bra{\Psi}$ and $\rho^{(E)}=\Tr_A\ket{\Psi}\bra{\Psi}$, respectively.
Their number equals $N=\text{min}(\text{dim}\,\BH_A,\text{dim}\,\BH_E)$.

The microscopic Born rule \eqref{sa1} follows from the orthogonality of $\ket{\tilde\psi_i}$:
\EQ{
\bra{\psi_i}\otimes\bra{\tilde\psi_i}\big({\cal O}_n\otimes 1_E\big)\ket{\psi_j}\otimes\ket{\tilde\psi_j}=
\bra{\psi_i}{\cal O}_n\ket{\psi_j}\bra{\tilde\psi_i}\tilde\psi_j\rangle=\delta_{ij}\bra{\psi_i}{\cal O}_n\ket{\psi_i}\ ,
}
which implies \eqref{sa1}.
The non-degeneracy condition rules out having more than one quantum microstate associated to each $\ket{\psi_i}$ since having two potential microstates $\ket{\psi_i}\otimes\ket{\psi'_i}$ and $\ket{\psi_i}\otimes\ket{\psi''_i}$ violates the non-degeneracy condition \eqref{sa2}.

This example forms the basis of the reduced density operator formalism developed in \cite{Hollowood:2013cbr,Hollowood:2013xfa,Hollowood:2013bja}.

\vspace{0.2cm}
\noindent{\bf Example 3:}
Consider the case of a small finite dimensional set of non-commuting observables $\EA$ in a large Hilbert space in a generic state $\ket{\Psi}$. We will assume that $\EA$ forms a Lie algebra, or can be completed into such an algebra. To be more precise, we will consider the case that the observables generate a $u(N)$ algebra. This example is not supposed to be realistic but it just illustrates how the formalism can cope with a non-abelian set of operators.

The complexified generators can be labelled by a pair $i,j\in\{1,2,\ldots,N\}$ with\footnote{The Hermitian combinations are ${\cal O}_{ii}$, ${\cal O}_{ij}+{\cal O}_{ji}$ and $i({\cal O}_{ij}-{\cal O}_{ji})$. The complexified operators satisfy an algebra under multiplication as well as commutation.}
\EQ{
[{\cal O}_{ij},{\cal O}_{kl}]=\delta_{jk}{\cal O}_{il}-\delta_{il}{\cal O}_{jk}\ .
}
We are assuming that the dimension of the algebra $N^2$ is much smaller than the dimension of the Hilbert space $\BH$. This means that a typical state $\ket{\Psi}$ is not annihilated by any element of $\EA$:
\EQ{
{\cal O}\ket{\Psi}\neq0\ ,\qquad \forall\ {\cal O}\in\EA\ .
}
In such a situation, the $N^2$ dimensional space $\BH_{\EA,\Psi}\subset\BH$ spanned by the independent states 
\EQ{
{\cal O}_{ij}\ket{\Psi}\ ,\qquad\forall\ i,j
}
carry an adjoint representation of the Lie algebra. A basis for this space consists of the orthonormal states $\ket{ij}$ labelled by a pair $i,j$ with
\EQ{
{\cal O}_{ij}\ket{kl}=\delta_{jk}\ket{il}\ .
}

The generators and states can always redefined by a unitary $\text{U}(N)$ transformation:
\EQ{
{\cal O}_{ij}\to U{\cal O}_{ij}U^{-1}\ ,\qquad \ket{ij}\to U\ket{ij}\ .
}
Given the typical state $\ket{\Psi}$, we 
can always use this freedom to perform a $\text{U}(N)$ transformation to align the algebra with the state $\ket{\Psi}$ in the sense that
\EQ{
\ket{\Psi}=\sum_{i=1}^Nc_i\ket{ii}+c_{N+1}\ket{\Psi^\perp}\ .
}
In this expression, $\ket{\Psi^\perp}$ is the component of $\ket{\Psi}$ lying orthogonal to $\BH_{\EA,\Psi}$.

The expression above identifies the $N+1$ quantum microstates as 
\EQ{
\ket{\Psi_i}=\ket{ii}\ ,\qquad i=1,2,\ldots,N\ ,\qquad\ket{\Psi_{N+1}}=\ket{\Psi^\perp}\ .
}
The quantum microstates are precisely the eigenstates of the Cartan subalgebra generated by the mutually-commuting operators ${\cal O}_{ii}$, $i=1,2,\ldots,N$ and, in particular, $\ket{\Psi_{N+1}}$ has zero eigenvalues for all the Cartan generators:
\EQ{
{\cal O}_{ii}\ket{\Psi_j}=\delta_{ij}\ket{\Psi_j}\ ,\qquad j=1,2,\ldots,N\ ,\qquad{\cal O}_{ii}\ket{\Psi_{N+1}}=0\ .
}
Note that the Cartan generators are the only operators with non-vanishing expectation values
\EQ{
\bra{\Psi}{\cal O}_{ij}\ket{\Psi}=|c_i|^2\delta_{ij}\ . 
}

\section{Time Dependent Observables}\label{a2}

In this appendix, we discuss how to general the formalism to the case of explicitly time dependent observables. The transition rates in \eqref{trt10} as generalized to 
\EQ{
T_{ij}=\text{max}\Big(-\frac2\hbar\IM\Big[\frac{c_i}{c_j}\bra{\Psi_j}\Big(H-i\hbar\frac d{dt}\Big)\ket{\Psi_i}\Big],0\Big)\ .
\label{trt9}
}

The discussion of locality in section \ref{s5} is modified in the following way, without any of the conclusions being changed. First of all, \eqref{pp2} generalizes to
\EQ{
\bra{\Psi_i}1_A\otimes \Big(H_E-i\hbar\frac d{dt}\Big)\ket{\Psi_j}=0\ ,\qquad i\neq j\ ,
}
where the derivative term here acts only on $\ket{\psi_{ja}}$. The final local expression for the transition rates is
\EQ{
T_{ij}=\text{max}\Big(-\frac2\hbar\IM\Big[\frac{c_i}{c_j}
\bra{\Psi_j}\Big\{\Big(H_A-i\hbar\frac d{dt}\Big)\otimes 1_E+H_\text{int}\Big\}\ket{\Psi_i}\Big],0\Big)\ ,
\label{pt3}
}
which generalizes \eqref{trt3}.

\end{document}